\documentclass[
  aps,
  amsmath,amssymb,reprint,superscriptaddress
]{revtex4-2}

\usepackage[dvipdfmx]{graphicx}

\usepackage{dcolumn}
\usepackage{bm}

\usepackage{braket}
\usepackage{color}
\usepackage{ulem}
\usepackage{comment}
\usepackage{here}

\begin{document}
\title{Four-body interactions in Kerr parametric oscillator circuits}
\author{Yohei Kawakami}
  \email{Contact author: yohei-kawakami@nec.com}
\affiliation{Secure System Platform Research Laboratories,
NEC Corporation, Kawasaki, Kanagawa 211-0011, Japan}
\affiliation{NEC-AIST Quantum Technology Cooperative Research Laboratory,
National Institute of Advanced Industrial Science and Technology (AIST),
Tsukuba, Ibaraki 305-8568, Japan
}
\author{Tomohiro Yamaji}
\affiliation{Secure System Platform Research Laboratories,
NEC Corporation, Kawasaki, Kanagawa 211-0011, Japan}
\affiliation{NEC-AIST Quantum Technology Cooperative Research Laboratory,
National Institute of Advanced Industrial Science and Technology (AIST),
Tsukuba, Ibaraki 305-8568, Japan
}
\author{Aiko Yamaguchi}
\affiliation{Secure System Platform Research Laboratories,
NEC Corporation, Kawasaki, Kanagawa 211-0011, Japan}
\affiliation{NEC-AIST Quantum Technology Cooperative Research Laboratory,
National Institute of Advanced Industrial Science and Technology (AIST),
Tsukuba, Ibaraki 305-8568, Japan
}
\author{Yuya Kano}
\affiliation{Secure System Platform Research Laboratories,
NEC Corporation, Kawasaki, Kanagawa 211-0011, Japan}
\affiliation{NEC-AIST Quantum Technology Cooperative Research Laboratory,
National Institute of Advanced Industrial Science and Technology (AIST),
Tsukuba, Ibaraki 305-8568, Japan
}
\author{\\Takaaki Aoki}
\affiliation{Global Research and Development Center for Business by Quantum-AI Technology (G-QuAT),\\
National Institute of Advanced Industrial Science and Technology (AIST), Tsukuba, Ibaraki 305-8568, Japan}
\author{Aree Taguchi}
\affiliation{NEC-AIST Quantum Technology Cooperative Research Laboratory,
National Institute of Advanced Industrial Science and Technology (AIST),
Tsukuba, Ibaraki 305-8568, Japan
}
\affiliation{Division of Physics, Faculty of Pure and Applied Sciences,\\
University of Tsukuba, Tsukuba, Ibaraki 305-8571, Japan}
\author{Kiyotaka Endo}
\affiliation{Secure System Platform Research Laboratories,
NEC Corporation, Kawasaki, Kanagawa 211-0011, Japan}
\affiliation{NEC-AIST Quantum Technology Cooperative Research Laboratory,
National Institute of Advanced Industrial Science and Technology (AIST),
Tsukuba, Ibaraki 305-8568, Japan
}
\author{Tetsuro Satoh}
\affiliation{Global Research and Development Center for Business by Quantum-AI Technology (G-QuAT),\\
National Institute of Advanced Industrial Science and Technology (AIST), Tsukuba, Ibaraki 305-8568, Japan}
\author{Ayuka Morioka}
\affiliation{Secure System Platform Research Laboratories,
NEC Corporation, Kawasaki, Kanagawa 211-0011, Japan}
\affiliation{NEC-AIST Quantum Technology Cooperative Research Laboratory,
National Institute of Advanced Industrial Science and Technology (AIST),
Tsukuba, Ibaraki 305-8568, Japan
}
\author{Yuichi Igarashi}
\affiliation{Secure System Platform Research Laboratories,
NEC Corporation, Kawasaki, Kanagawa 211-0011, Japan}
\affiliation{NEC-AIST Quantum Technology Cooperative Research Laboratory,
National Institute of Advanced Industrial Science and Technology (AIST),
Tsukuba, Ibaraki 305-8568, Japan
}
\author{\\Masayuki Shirane}
\affiliation{Secure System Platform Research Laboratories,
NEC Corporation, Kawasaki, Kanagawa 211-0011, Japan}
\affiliation{NEC-AIST Quantum Technology Cooperative Research Laboratory,
National Institute of Advanced Industrial Science and Technology (AIST),
Tsukuba, Ibaraki 305-8568, Japan
}
\author{Tsuyoshi Yamamoto}
  \email{Contact author: tsuyoshi.yamamoto@nec.com}
\affiliation{Secure System Platform Research Laboratories,
NEC Corporation, Kawasaki, Kanagawa 211-0011, Japan}
\affiliation{Global Research and Development Center for Business by Quantum-AI Technology (G-QuAT),\\
National Institute of Advanced Industrial Science and Technology (AIST), Tsukuba, Ibaraki 305-8568, Japan}
\affiliation{Division of Physics, Faculty of Pure and Applied Sciences,\\
University of Tsukuba, Tsukuba, Ibaraki 305-8571, Japan}
\begin{abstract}
  We theoretically present new unit circuits of Kerr parametric oscillators (KPOs) with four-body interactions,
  which enable the scalable embedding of all-to-all connected logical Ising spins using the Lechner-Hauke-Zoller (LHZ) scheme.
  These unit circuits enable four-body interactions using linear couplers,
  making the circuit fabrication and characterization much simpler than those of conventional unit circuits with nonlinear couplers.
  Numerical calculations indicate that the magnitudes of the coupling constants can be comparable to those in conventional circuits.
  On the basis of this theory, we designed a four-KPO circuit and experimentally confirmed the four-body correlation
  by measuring the pump-phase dependence of the parity of the four-KPO states.
  We show that the choice of the pump frequencies are important
  not only to enable the four-body interaction,
  but to cancel the effects of other unwanted interactions.
  Using the circuit, we demonstrated the quantum annealing based on the LHZ scheme,
  where the strength of the interaction between the logical Ising spins
  is mapped to the local field and controlled by a coherent drive applied to each KPO.
\end{abstract}
\maketitle
\section{Introduction}

Kerr parametric oscillators (KPOs) are promising as components for quantum-information processing.
~\cite{Puri:2017jmb,Goto:2016avj,Puri:2019jjg,Gautier:2021jjv,Puri:2020vxr,You:2019qlr,andersen2020quantum,Darmawan:2021kwi,Mirrahimi:2014fhp,Kanao:2021lzl,Chono:2022zsh}.
KPOs have been studied using flux-driven SQUIDs (superconducting quantum interference devices)
~\cite{Iyama:2023wlx,Kwon:2022xaf,Yamaji:2020pfg,Yamaji:2022ikh,Yamaji:2025vth,Yamaguchi:2023jrp, Aoki:2023zuj, Masuda:2024djh, Aoki:2024fsg}
and charge-driven SNAILs (superconducting nonlinear asymmetric inductive elements)
~\cite{Frattini:2022dxx,Venkatraman:2022ibc,Hajr:2024eme,Chavez-Carlos:2022qqj,Venkatraman:2022dxu,Miano:2021gmr,He:2023ejn,Grimm:2020kfs,Qing:2024xfv}.
These devices have been extensively investigated, for example, for applications in gate-based qubits (hence, they are often called Kerr cat qubits). 
Consideration is starting to be given to the utilization of Josephson-junction components
other than SQUIDs and SNAILs~\cite{Bhandari:2024kdo,Hua:2024xhq}.

Quantum annealing is also a widely recognized application of KPOs,
where their degenerate coherent states (oscillation states) correspond to a binary Ising spin~\cite{Goto:2016owt,Goto:2019kqy,Goto:2020ptv, Goto:boltzman}.
Although this is anticipated to solve combinatorial optimization problems,
implementing long-range two-spin interactions in Ising Hamiltonians
in practical devices presents significant challenges.

Studies~\cite{Sourlas:2005,Lechner:2015bal} have indicated that 
four-body interactions between nearest-neighboring physical spins
can be used to embed the all-to-all two-body interactions
of logical Ising spins
in a physical spin network.
This approach is known as parity encoding or
the Lechner-Hauke-Zoller scheme (the LHZ scheme).
A unit circuit using KPOs with a four-body interaction between them has been proposed~\cite{Puri:2017hqg}
and theoretically investigated~\cite{Kanao:2021bxu,Miyazaki:2025gch,Matsuzaki:2025nhd},
where the KPOs are connected to a nonlinear coupler, the Kerr nonlinearity of which mediates the four-body interaction.
However, experimental implementations of the unit circuit have not yet been reported.

Here, we theoretically investigate unit circuits using KPOs with four-body interactions.
Specifically, we study four-body interactions originating from Kerr nonlinearities of KPOs.
We propose various unit circuits on the basis of this investigation.
These circuits are anticipated to significantly simplify the physical implementation of the Ising machine
in terms of fabrication and operation.
We also experimentally demonstrate the four-body interactions using the new unit circuits.

This article is organized as follows.
In Sec.~\ref{sec:recap}, we recap the conventional unit circuit~\cite{Puri:2017hqg} and the four-body interaction originating from a coupler nonlinearity.
In Sec.~\ref{sec:four}, we introduce new unit circuits in which the four-body interaction arises from KPO nonlinearities.
In Sec.~\ref{sec:locality}, we investigate the locality of the four-body interactions and scalability of our unit circuits.
In Sec.~\ref{sec:numerical}, we present our numerical study on the magnitudes of the four-body interactions.
In Sec.~\ref{sec:exp}, we present our experimental investigation on the four-body interaction in one of our unit circuits using superconducting KPOs.
Finally, we conclude the article in Sec.~\ref{sec:conclusion}.

\section{Recap of conventional unit circuit} \label{sec:recap}
\begin{figure}[tb]
\includegraphics[width=8cm]{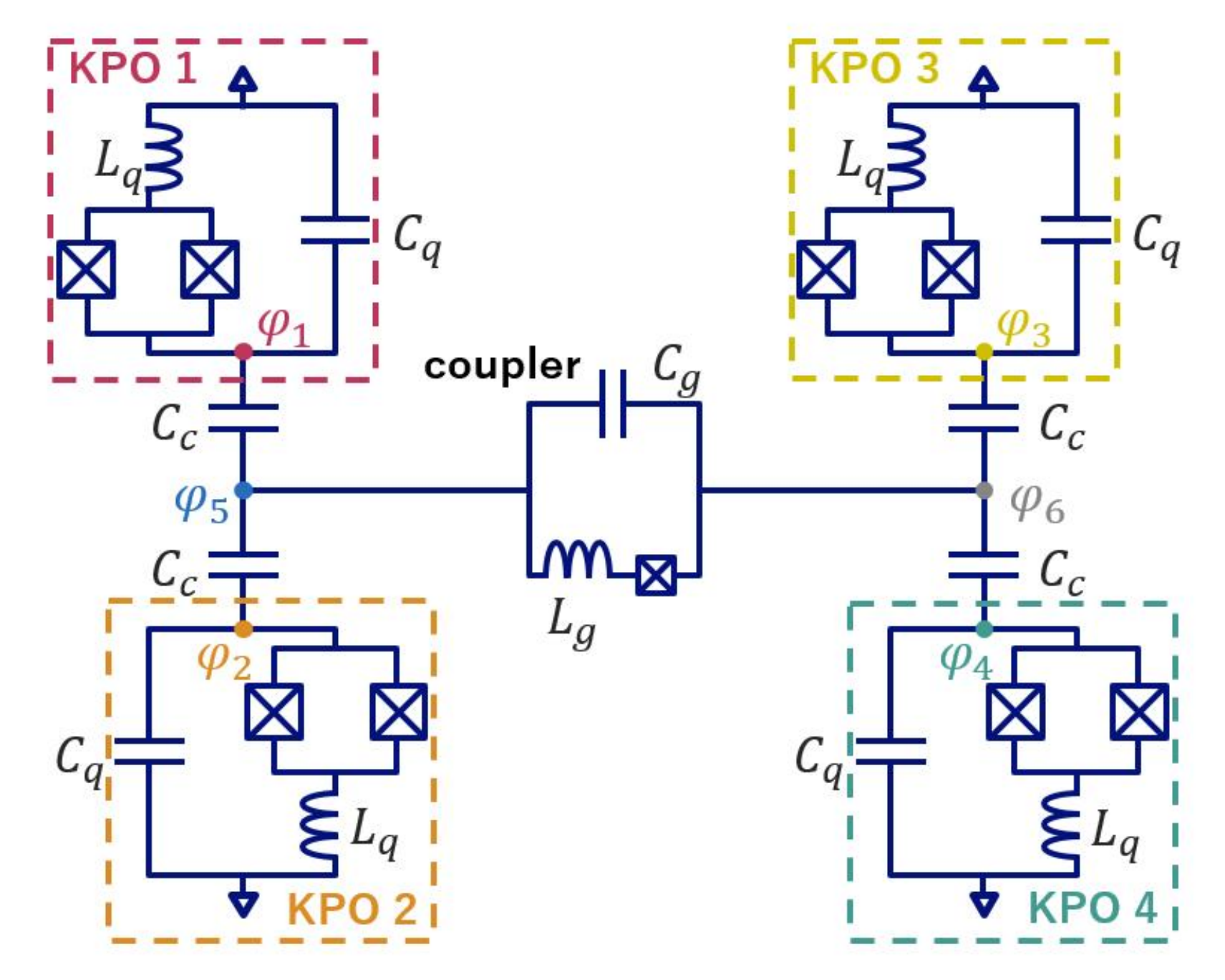}
\caption{\label{fig:original}
Schematic of a conventional unit circuit with a four-body interaction between four KPOs.
We label the KPOs as KPO 1 to KPO 4.
A box with a cross mark inside represents a Josephson junction.
The $\varphi_k$ ($k=1,...,6$) is the reduced node flux,
$C_q$s and $C_g$ are the cavity capacitances of the KPOs and coupler,
respectively, $C_c$s are the coupling capacitances,
and $L_q$s and $L_g$ are the linear inductances of the KPOs and coupler, respectively.
}
\end{figure}

Puri et al. proposed a unit circuit of an Ising machine with a four-body interaction between KPOs~\cite{Puri:2017hqg}.
Figure~\ref{fig:original} illustrates a circuit using KPOs based on their unit circuit
comprising four KPOs and a nonlinear coupler.
The $\varphi_k$ ($k=1,...,6$) is the reduced node flux~\cite{Vool:2017dhh},
$C_q$s and $C_g$ are the cavity capacitances of the KPOs and coupler,
respectively, and $C_c$s are the coupling capacitances.
The $C_g$ is added to the circuit to adjust the coupler resonance frequency to be close to the KPO resonance frequencies.
Each KPO is modeled as a capacitively-shunted SQUID
following Ref.~\cite{Puri:2017hqg},
but we added a linear inductance $L_q$ in series to the SQUID~\cite{Yamaguchi:2023jrp}.
The linear inductance $L_g$ is also connected in series with the single junction in the coupler.

The Hamiltonian of the unit circuit is obtained as follows
(see Appendix A for the derivation):
\begin{align}
  H/\hbar &= \sum_j^4 \left\{ \omega_j a_j^\dagger a_j -\frac{K_j}{12} \left( a_j +a_j^\dagger \right)^4 \right\} \notag \\
  &+\sum_j^4 p_j \left( a_j +a_j^\dagger\right)^2 \cos \omega_{\mathrm{p}j} t \notag \\
  &+\omega_g a_g^\dagger a_g -\frac{K_g}{12} \left( a_g +a_g^\dagger \right)^4 \notag \\
  &-\sum_j^4 s_j g_{j} \left( a_j -a_j^\dagger \right) \left( a_g -a_g^\dagger \right), \label{eq:original_hamiltonian}
\end{align}
where $j$ ($j = 1, ..., 4$) represents the label of the KPOs, and
$\omega_j$, $K_j$, and $a_j$ are the resonance frequency, Kerr nonlinearity, and annihilation operator of the fundamental mode of KPO $j$, respectively.
Similarly, $\omega_g$, $K_g$, and $a_g$ are the resonance frequency, Kerr nonlinearity, and annihilation operator of the coupler.
Here, $K_j$s and $K_g$ are assumed to be positive in this notation.
The second term is the pump drive (two-photon drive) term for the KPOs,
where $p_j$ and $\omega_{\mathrm{p}j}$ ($\approx 2\omega_j$) represent the magnitude and frequency of the pump drive for KPO $j$, respectively.
It should be noted that direct two-body couplings between the KPOs are omitted here and will be introduced in the following section.

The $g_{j}$ represents the two-body coupling constant between KPO $j$ and the coupler, and
$s_j$ is a factor related to the sign of the coupling, $s_j = 1$ for $j=1, 2$ and $s_j = -1$ for $j=3, 4$.
To exclude the two-body interactions at the leading order, we transform the Hamiltonian using the following unitary operator~\cite{Blais:2020wjs} (see Appendix B for details):
\begin{equation}
  U_g = {\rm exp} \sum_j^4 \left\{ -s_j\tilde{g}_{j} \left( a_j^\dagger a_g -a_j a_g^\dagger \right)\right\},
\end{equation}
\begin{equation}
  \tilde{g}_{j} = \frac{g_{j}}{\Delta_{j}}, \qquad \Delta_{j} = \omega_j -\omega_g,
\end{equation}
where $\tilde{g}_j$ is the perturbative expansion parameter ($|\tilde{g}_j| \ll 1$).
With this transformation, we obtain four-body-interaction terms from the coupler nonlinearity as
\begin{align}
  & U_g^\dagger \left( -\frac{K_g}{2} a_g^{\dagger 2} a_g^2 \right) U_g \notag \\
  & \to -g^{(4)} \left( a_1^\dagger a_2^\dagger a_3 a_4 +a_1^\dagger a_2 a_3^\dagger a_4 +a_1^\dagger a_2 a_3 a_4^\dagger +{\rm h.c.}\right),
\end{align}
\begin{equation}
  g^{(4)} = 2 \tilde{g}_{1} \tilde{g}_{2} \tilde{g}_{3} \tilde{g}_{4} K_g. \label{eq:coupler_4bi}
\end{equation}
The magnitude of $g^{(4)}$ is suppressed by the quartic factor $\tilde{g}_{1} \tilde{g}_{2} \tilde{g}_{3} \tilde{g}_{4}$.
To make $g^{(4)}$ large, we need to use a coupler with a large Kerr nonlinearity such as a transmon~\cite{Koch:2007hay} or quarton~\cite{Yan:2020vjc}, etc.
since $g^{(4)}$ is proportional to $K_g$.

By driving the KPOs with pump frequencies $\omega_{\mathrm{p}j}$
through inductive couplings between drive lines and SQUIDs,
they undergo parametric oscillation at $\omega_{\mathrm{p}j}/2$.
We obtain the Hamiltonian in the rotating frame at $\omega_{\mathrm{p}j}/2$ using the following unitary transformation:
\begin{equation}
  U_{\omega_{\mathrm{p}}/2} = {\rm exp} \left( -i\sum_j^4 \frac{\omega_{\mathrm{p}j} t}{2} a_j^\dagger a_j \right).
\end{equation}
Under this transformation, the annihilation operator $a_j$ transforms as
\begin{equation}
  U_{\omega_{\mathrm{p}}/2}^\dagger a_j U_{\omega_{\mathrm{p}}/2} = e^{-i\omega_{\mathrm{p}j} t/2} a_j\label{eq:rf-trans}.
\end{equation}
By adopting the four-body mixing condition $\omega_{\mathrm{p}1} +\omega_{\mathrm{p}2} = \omega_{\mathrm{p}3} +\omega_{\mathrm{p}4}$,
we obtain only the following terms under the rotating-wave approximation:
\begin{equation}
  -g^{(4)} \left( a_1^\dagger a_2^\dagger a_3 a_4 +{\rm h.c.} \right).
\end{equation}
Note that other terms of the four-body interaction remain
when a different condition, such as $\omega_{\mathrm{p}1} +\omega_{\mathrm{p}3} = \omega_{\mathrm{p}2} +\omega_{\mathrm{p}4}$ or $\omega_{\mathrm{p}1} +\omega_{\mathrm{p}4} = \omega_{\mathrm{p}2} +\omega_{\mathrm{p}3}$, is adopted.

\section{Four-body interactions originating from KPO nonlinearities} \label{sec:four}
In the previous section, we reviewed the four-body interaction originating from a coupler nonlinearity.
In this section, we show that four-body interactions can also arise from another source: Kerr nonlinearities of KPOs.

\subsection{Basic formulation}
Although not included in Eq.~(\ref{eq:original_hamiltonian}),
the following two-body interactions between the KPOs exist in the circuit shown in Fig.~\ref{fig:original}:
\begin{equation}
  -\sum_{j < k}^4 h_{jk} \left( a_j -a_j^\dagger\right) \left( a_k -a_k^\dagger \right),
\end{equation}
where $h_{jk}$ is the two-body coupling constant between KPOs $j$ and $k$.
The details of the derivation are provided in Appendix A.
Similar to the previous section, we transform the Hamiltonian using the following unitary operator:
\begin{equation}
  U_q = {\rm exp} \sum_{j<k}^4 \left\{ -\tilde{h}_{jk} \left( a_j^\dagger a_k -a_j a_k^\dagger \right)\right\}, \label{eq:Uq}
\end{equation}
\begin{equation}
  \tilde{h}_{jk} = \frac{h_{jk}}{\Delta_{jk}}, \qquad \Delta_{jk} = \omega_j -\omega_k,
\end{equation}
where $h_{jk} = h_{kj}$ and $\Delta_{jk} = -\Delta_{kj}$.
We thus obtain a four-body interaction from the KPO nonlinearities:
\begin{align}
  & U_q^\dagger \left( -\sum_j^4 \frac{K_j}{2} a_j^{\dagger 2} a_j^2 \right) U_q \notag \\
  & \to -h^{(4)} \left( a_1^\dagger a_2^\dagger a_3 a_4 +a_1^\dagger a_2 a_3^\dagger a_4 +a_1^\dagger a_2 a_3 a_4^\dagger +{\rm h.c.}\right),
\end{align}
\begin{equation}
  h^{(4)} = \sum_{j\neq k \neq l \neq m}^4 2 \tilde{h}_{kj} \tilde{h}_{lj} \tilde{h}_{mj} K_j. \label{eq:original_h4}
\end{equation}
This is another mechanism for generating a four-body interaction in the circuit.
It should be noted that the new coupling constant $h^{(4)}$ is derived from the third order of perturbation,
while $g^{(4)}$ is from the fourth order.

The coupling constant is explicitly expressed as
\begin{align}
  h^{(4)} &= 2K_1 \tilde{h}_{21}\tilde{h}_{31}\tilde{h}_{41} +2K_2 \tilde{h}_{12}\tilde{h}_{32}\tilde{h}_{42} \notag \\
  &+2K_3 \tilde{h}_{13}\tilde{h}_{23}\tilde{h}_{43} +2K_4 \tilde{h}_{14}\tilde{h}_{24}\tilde{h}_{34} \label{eq:original-h4}.
\end{align}
To discuss the coupling constant in detail,
we make approximations for $h_{jk}$s.
First, since the variations in $\omega_j$s controlled by an external flux considered in this article are much smaller than $\omega_j$s,
we neglect the frequency dependence of $h_{jk}$.
Second, with the geometric symmetry of the circuit in Fig.~\ref{fig:original},
we assume the following conditions for $h_{jk}$s:
\begin{equation}
  h_{12} = h_{34}, \qquad h_{13} = h_{14} = h_{23} = h_{24}.
\end{equation}
Third, we impose that the frequency detuning
$\delta_j$ ($= \omega_j -\omega_{\mathrm{p}j}/2$) is identical for all $j$ for simplicity. 
We thus obtain the following condition on $\omega_j$s:
\begin{equation}
  \omega_1 +\omega_2 = \omega_3 +\omega_4\label{eq:freq_condition}.
\end{equation}
Under these approximations, we obtain the following formula for $h^{(4)}$:
\begin{equation}
  h^{(4)} \simeq 2h_{12}h_{13}^2 \frac{\left( K_2 -K_1\right)\Delta_{34} +\left( K_3 -K_4\right)\Delta_{12}}{\Delta_{12}\Delta_{13}\Delta_{14}\Delta_{34}}. \label{eq:h4}
\end{equation}
The $h^{(4)}$ is symmetric with respect to the replacement of indices $1 \leftrightarrow 2$ and/or $3 \leftrightarrow 4$, which is consistent with the symmetry of the circuit geometry in Fig.~\ref{fig:original}.
Therefore, we need to make $K_1\neq K_2$ and/or $K_3 \neq K_4$ to obtain a nonzero $h^{(4)}$.
Its magnitude can be enhanced by making the signs of
$K_2$ and $K_3$ different from those of $K_1$ and $K_4$.
SNAILs, for example, 
can exhibit both positive and negative nonlinearities~\cite{Frattini:2017vvt,Frattini:2018ahg}.
We investigate the $h^{(4)}$ of such a circuit in Sec.\ref{sec:numerical}.

\begin{figure*}[]
\includegraphics[width=16cm]{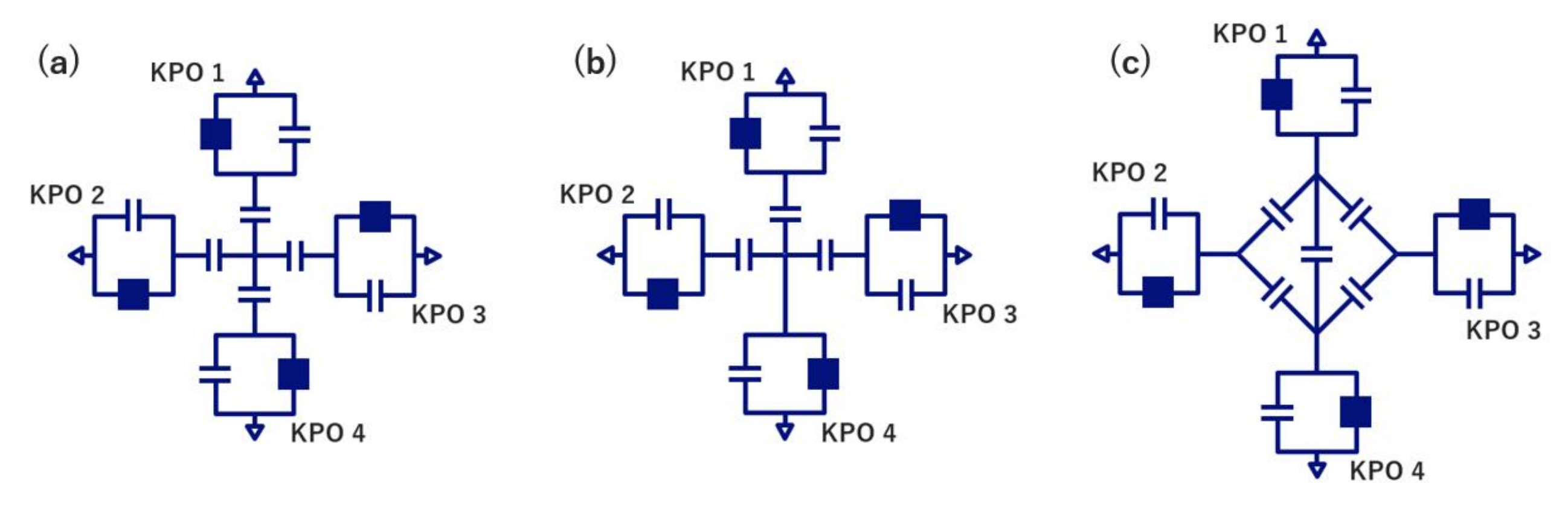}
\caption{\label{fig:circuits_qubit_4bi}
Unit circuits with four-body interactions originating from KPO nonlinearities.
The black boxes represent inductive components of the KPOs.
In (a), four KPOs are capacitively coupled to each other without going through any grounded elements.
In (b), KPOs 1 and 2, 1 and 3, and 2 and 3 are coupled via KPO 4,
resulting in the suppression of $h_{12}$, $h_{13}$, and $h_{23}$.
In (c), KPOs 2 and 3 are coupled via KPOs 1 and 4,
resulting in the suppression of $h_{23}$.
}
\end{figure*}

It is worth mentioning that the formula in Eq.~(\ref{eq:h4}) does not include any degrees of freedom of the coupler.
Therefore, we can use a simpler circuit that does not include a coupler, as shown in Fig.~\ref{fig:circuits_qubit_4bi}(a).
This circuit is much more favorable for the realization of the Ising machine
in terms of
the simplicity of fabrication and operation.
Moreover, there is no concern about the excitation of the coupler mode~\cite{Puri:2017hqg}. 

\subsection{Other unit circuits}
In $h^{(4)}$, which was derived in the previous subsection,
all $K_j$s participate in the coupling.
Alternatively, we can eliminate some of the $K_j$s from $h^{(4)}$ by changing the layout of capacitances of the unit circuit.

Figure \ref{fig:circuits_qubit_4bi}(b) shows the case where the coupling capacitance for KPO 4 is removed.
In this case, KPOs 1 and 2, for example, are connected via KPO 4, which is grounded.
Therefore, the coupling constants $h_{12}$, $h_{13}$, and $h_{23}$ are suppressed compared with $h_{14}$, $h_{24}$, and $h_{34}$
(more details are given in the next section).
Assuming the same conditions as in the previous subsection,
we have the following four-body coupling:
\begin{align}
  \tilde{h}^{(4)} &= 2 \tilde{h}_{14}\tilde{h}_{24}\tilde{h}_{34} K_4 \notag \\
  &\simeq -2 {h^{\prime}}^3 \frac{K_4}{\Delta_{13} \Delta_{14} \Delta_{34}}, \label{eq:one_origination}
\end{align}
where $h^{\prime} \equiv h_{14} = h_{24} = h_{34}$.
Only $K_4$ contributes to the four-body coupling $\tilde{h}^{(4)}$.
While $h^{(4)}$ is zero when $K_j$s are all identical,
$\tilde{h}^{(4)}$ is always nonzero.
We can thus use KPOs with identical $K_j$s in the unit circuit shown in Fig.~\ref{fig:circuits_qubit_4bi}(b).
This may simplify the design and fabrication of chips.

By changing the geometry of the capacitances,
we can make $K_1$ contribute to the four-body coupling in addition to $K_4$.
The circuit is shown in Fig.~\ref{fig:circuits_qubit_4bi}(c).
The $h_{23}$ is suppressed because 
KPOs 2 and 3 are coupled via KPOs 1 and 4, which are grounded.
We thus obtain the following formula for the four-body coupling:
\begin{align}
  \tilde{\tilde{h}}^{(4)} &= K_1 \tilde{h}_{21}\tilde{h}_{31}\tilde{h}_{41} +K_4 \tilde{h}_{14}\tilde{h}_{24}\tilde{h}_{34} \notag \\
  &\simeq -2{h^{\prime\prime}}^3 \frac{K_1 \Delta_{34} +K_4 \Delta_{12}}{\Delta_{12}\Delta_{13}\Delta_{14}\Delta_{34}}, \label{eq:two_origination}
\end{align}
where we represent the two-body couplings other than $h_{23}$ as $h^{\prime\prime}$.
Since both $K_1$ and $K_4$ contribute to the coupling,
we can obtain a larger magnitude compared with $\tilde{h}^{(4)}$.
We can also use KPOs with identical $K_j$s.
However, this circuit may not be practical as a unit cell,
as implementing it in a large-scale system would require dozens of distinct pump frequencies
as described later in Sec. \ref{sec:locality}.

\section{Locality and scalability} \label{sec:locality}
\begin{figure}[tb]
\includegraphics[width=8cm]{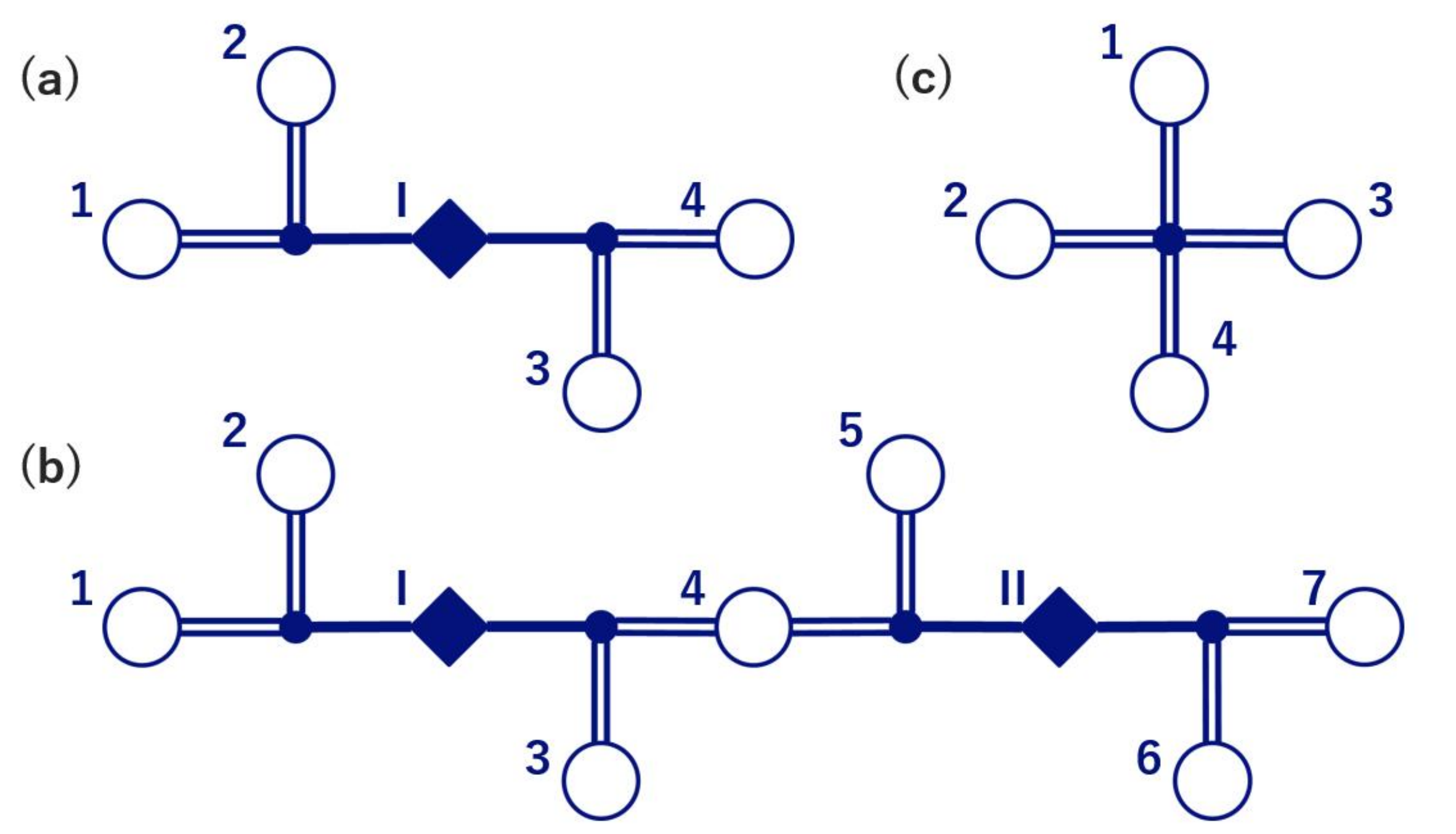}
\caption{\label{fig:net_diff} Abstractions of capacitive components.
The white circles, navy diamonds, and navy circles represent grounded KPOs, ungrounded couplers, and nodes (terminals of connections), respectively.
The double lines indicate connecting wires interrupted by a coupling capacitance $C_c$.
(a) Unit circuit with an ungrounded coupler.
(b) Extension of (a). There are two ungrounded couplers and seven KPOs.
(c) Unit circuit without a coupler.
}
\end{figure}

Up to this point, we have discussed four-body interactions within the unit structure.
To scale this structure up into a circuit usable for quantum computation,
it is necessary that the four-body interactions are local and that the unit structure itself is scalable.
In this section, to consider larger structures, we will first re-illustrate the circuit in a simplified manner and then discuss the locality of interactions and the scalability of the circuit.
Figure~\ref{fig:net_diff}(a) shows the abstraction of the circuit shown in Fig.~\ref{fig:original}, with a focus on the capacitive components.
For simplicity, we assume that the KPOs and coupler have the same capacitances $C$
that are much larger compared with $C_c$, i.e., $C \gg C_c$.
By calculating the inverse of the capacitance matrix following Vool et al.~\cite{Vool:2017dhh} (see Appendix A for details),
we obtain the coupling constant $g_j$ between KPO $j$ and the coupler ($\propto C_c/C$), as well as
the coupling constant $h_{jk}$ between KPOs $j$ and $k$ ($\propto C_c/C$),
where we used $C +C_c \simeq C$.
Note that $h_{jk}$ is further suppressed by a factor $C_c/C$ compared with $g_{j}$ if the coupler is grounded.
This circuit is discussed further in Appendix D.

We extend the model to include seven KPOs and two couplers, as shown in Fig.~\ref{fig:net_diff}(b).
We consider the couplings between a KPO and coupler across the central KPO (KPO 4),
such as the coupling between KPO 1 and Coupler II.
In this case, the coupling constant is suppressed by a factor $C_c/C$ compared with that between a KPO and neighboring coupler.
For example,
$g_{\mathrm{1\scalebox{0.7}{II}}}$ is much smaller than $g_{\mathrm{1\scalebox{0.7}{I}}}$ since
$g_{\mathrm{1\scalebox{0.7}{II}}} \propto (C_c/C)^2$ while $g_{\mathrm{1\scalebox{0.7}{I}}} \propto (C_c/C)$, where $g_{\mathrm{1\scalebox{0.7}{I}}}$ ($g_{\mathrm{1\scalebox{0.7}{II}}}$) represents the two-body coupling between KPO 1 and Coupler I (II).
When a two-body interaction spans over $n$ grounded elements,
its coupling constant is suppressed by a factor of $(C_c/C)^n$.
Since $g^{(4)}$ includes $g_{j}$s,
the four-body interaction is also suppressed when it crosses grounded elements.
Therefore, we regard the four-body interaction as a local interaction (a short-range interaction) and it is scalable,
as interactions between KPOs that are not directly connected by a coupler are negligible at the leading order.

\begin{figure}[tb]
\includegraphics[width=8cm]{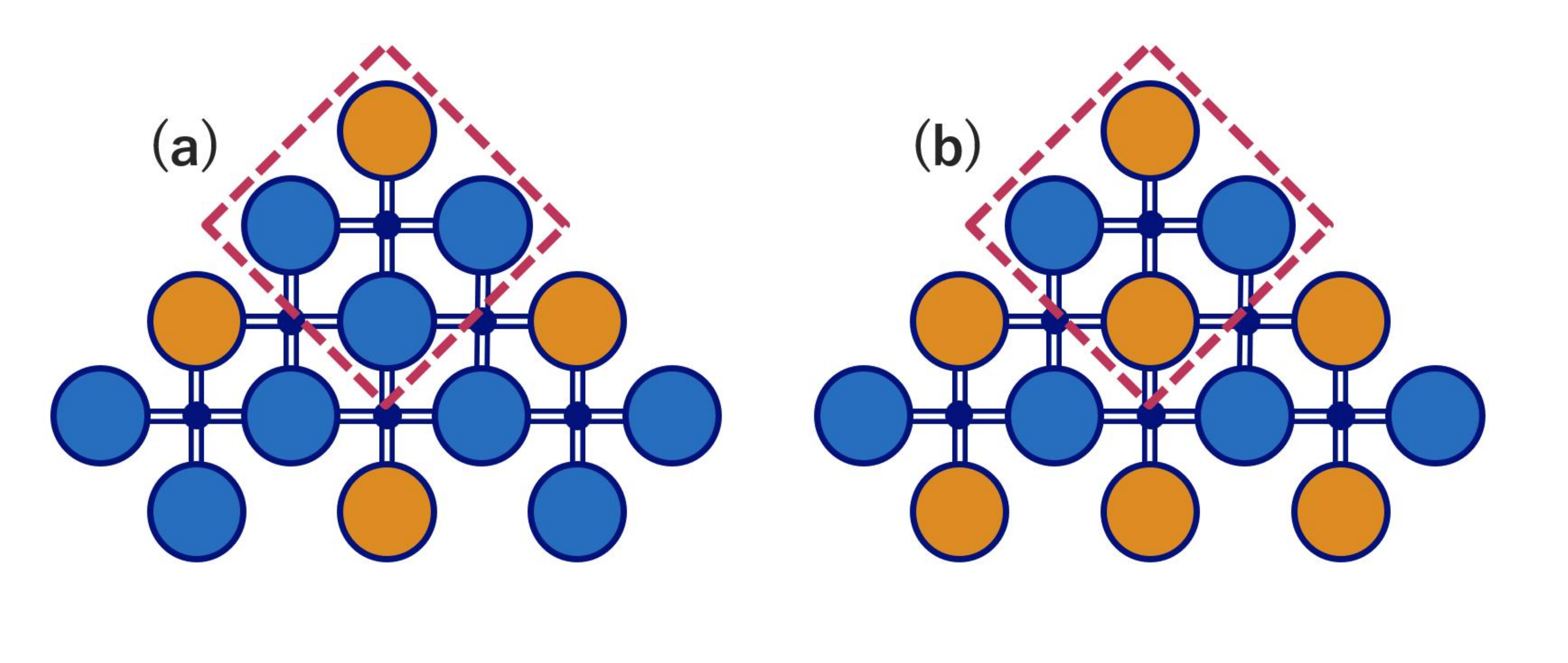}
\caption{\label{fig:scalability_diff_kerr} KPO arrangement based on the LHZ scheme~\cite{Sourlas:2005, Lechner:2015bal}
with $h^{(4)}$.
The regions outlined by the red dashed squares correspond to a unit cell.
In (a), there is one KPO (orange) that has a different Kerr nonlinearity from other three KPOs (blue) in each unit circuit.
In (b), there are two KPOs that have different Kerr nonlinearities from other two KPOs in each unit circuit.
}
\end{figure}

The $h^{(4)}$ is also a local interaction as $h_{jk}$s are similarly suppressed,
regardless of whether the circuit has couplers or not [the abstraction of the circuit without a coupler is shown in Fig.~\ref{fig:net_diff}(c)].
When using $h^{(4)}$, it is important to carefully consider the configuration of the Kerr nonlinearities of KPOs,
as certain combinations of them can lead to the cancellation of $h^{(4)}$ [see Eq.~(\ref{eq:h4})].
Examples of KPO arrangement based on the LHZ scheme~\cite{Sourlas:2005, Lechner:2015bal} are shown in Fig.~\ref{fig:scalability_diff_kerr}. 
The blue and orange circles represent KPOs with two different nonlinearities.
In Fig.~\ref{fig:scalability_diff_kerr}(a), one of the four KPOs in each unit cell (orange) has a different Kerr nonlinearity from the other three KPOs (blue)
(for example, $K_1$ in Eq.~(\ref{eq:h4}) can be assigned to orange and $K_2$, $K_3$, and $K_4$ to blue),
while in Fig.~\ref{fig:scalability_diff_kerr}(b) two of them do
($K_1$ and $K_4$ to orange, and $K_2$ and $K_3$ to blue).
In both cases, we can arrange KPOs to a large scale and implement a nonzero $h^{(4)}$ in each unit cell.

\begin{figure}[tb]
\includegraphics[width=8cm]{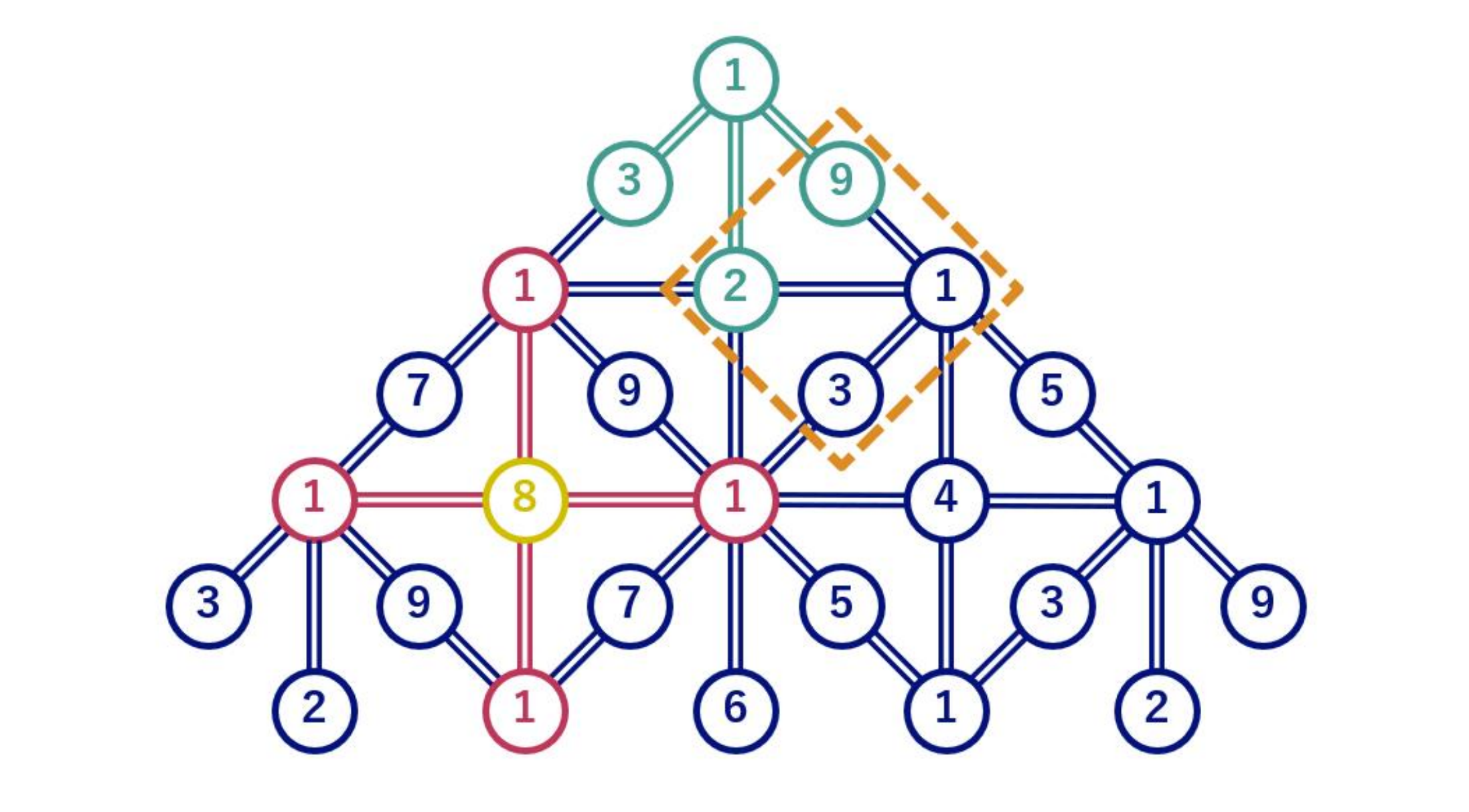}
\caption{\label{fig:scalability_limit} Large-scale network based on the unit circuit in Fig.~\ref{fig:circuits_qubit_4bi}(b),
where the four-body interaction originates from only one Kerr nonlinearity ($K_4$ in the circuit).
The subset of the green KPOs represents a unit cell as an example.
Note that the region outlined by the orange dashed square also represents a unit cell.
There is a four-body interaction between the red KPOs, which originates from the yellow KPO.
}
\end{figure}

Next, we examine the scalability of $\tilde{h}^{(4)}$, which is generated by the circuit shown in Fig.~\ref{fig:circuits_qubit_4bi}(b).
This circuit is also scalable.
In Fig.~\ref{fig:scalability_limit}, we represent
a larger-scale network based on a unit circuit of Fig.~\ref{fig:circuits_qubit_4bi}(b).
The numbers in the circles represent indices for pump frequencies.
We find that a four-body interaction can be implemented in each unit cell
by using nine distinct pump frequencies.
For example, we can impose the following four-body mixing conditions.
\begin{align}
  \omega_{\mathrm{p}1} +\omega_{\mathrm{p}2} &= \omega_{\mathrm{p}3} +\omega_{\mathrm{p}9}, \notag \\
  \omega_{\mathrm{p}1} +\omega_{\mathrm{p}8} &= \omega_{\mathrm{p}7} +\omega_{\mathrm{p}9}, \notag \\
  \omega_{\mathrm{p}1} +\omega_{\mathrm{p}4} &= \omega_{\mathrm{p}3} +\omega_{\mathrm{p}5}, \notag \\
  \omega_{\mathrm{p}1} +\omega_{\mathrm{p}6} &= \omega_{\mathrm{p}7} +\omega_{\mathrm{p}5},\label{eq:limit_conditions}
\end{align}
where $\omega_{\mathrm{p}j}$s ($j=1,..., 9$) are the pump frequencies.

The green KPOs in Fig.~\ref{fig:scalability_limit} form a unit cell that satisfies the condition of the first line in Eq.~(\ref{eq:limit_conditions}).
Note that a four-body interaction occurs also between the four KPOs with $\omega_{\mathrm{p}1}$ (shown in red) surrounding the KPO with $\omega_{\mathrm{p}8}$ (shown in yellow),
which acts as a coupler since these pump frequencies meet a four-body mixing condition.
Similar combinations of KPOs with $\omega_{\mathrm{p}1}$ around the KPOs with $\omega_{\mathrm{p}2}$ and $\omega_{\mathrm{p}4}$
can also be observed in the figure.
However, these four-body interactions are of the fourth order of perturbation and 
are therefore negligible compared with other third-order four-body interactions.

When extending the circuit shown in Fig.~\ref{fig:circuits_qubit_4bi}(c),
much more pump frequencies are required
to suppress unwanted four-body interactions
because the number of connections between KPOs increases
compared with the circuit of Fig.~\ref{fig:scalability_limit}.

\section{Numerical study of four-body couplings} \label{sec:numerical}
We now present numerical calculations for the coupling constants of various four-body interactions $g^{(4)}$, $h^{(4)}$, and $\tilde{h}^{(4)}$.
First, we reformulate these coupling constants for each relevant circuit.
For $h^{(4)}$, we consider two circuits.
The values of the parameters used in all circuits are summarized in Table.~\ref{tab:parameters}.
To compare four-body couplings for different circuits,
we adjust the coupling capacitances to ensure that the two-body couplings
are equal for all the circuits.
\begin{table*}
\caption{\label{tab:parameters}Parameters of the circuits for which the four-body coupling constants are calculated.
The $C_q$ is the cavity capacitance of the KPOs in the circuits shown in Figs.~\ref{fig:original}, \ref{fig:circuits_numerical}(a), and \ref{fig:circuits_numerical}(c).
The $L_q$ is the linear inductance of the KPOs,
and $C_c$s are the coupling capacitances in all circuits.
The $C_c$s are adjusted to set the two-body couplings $g_g/2\pi=h_q/2\pi=h_{\mathrm{\scalebox{0.7}{QN}}}/2\pi=5$ MHz.
The $C_g$s are the cavity capacitances and $L_g$ is the linear inductance of the coupler in Fig.~\ref{fig:original}.
In Fig.~\ref{fig:circuits_numerical}(a), $I_{0\mathrm{sr}}$ is the critical current of the junction in series with a SQUID in KPOs 1 and 4.
In Fig.~\ref{fig:circuits_numerical}(b), $C_{qj}$ is the cavity capacitance of KPO $j$.
The $I_{0\mathrm{sn}}$ is the critical current of the junctions in series in the SNAIL,
and $\gamma$ is the ratio of the critical currents of the Josephson junctions in the SNAIL.
}
\begin{ruledtabular}
\begin{tabular}{cccccc}
  $g^{(4)}$ in Fig.~\ref{fig:original} [Eq.~(\ref{eq:g4_num})] &$C_q$ [fF]& $L_q$ [pH] &$C_c$ [fF]&$C_g$ [fF] &$L_g$ [pH] \\
  KPO-like coupler&$500$&$100$&$1$& $500$ &$100$ \\
  transmon-like coupler&$500$&$100$&$1/\sqrt{5}$&$100$&$100$ \\ \hline
  $h^{(4)}$ in Fig.~\ref{fig:circuits_numerical}(a) [Eq.~(\ref{eq:approx_h4})] &$C_q$ [fF]& $L_q$ [pH] &$C_c$ [fF] & $I_{0\mathrm{sr}}$ [nA] & \\
  &$500$&$100$&$2$&$1500$ &\\ \hline
  $h^{(4)}$ in Fig.~\ref{fig:circuits_numerical}(b) [Eq.~(\ref{eq:h4_SNAIL})] & $C_{q1,4}/C_{q2,3}$ [fF] & $L_q$ [pH] & $C_c$ [fF] & $I_{0\mathrm{sn}}$ [nA] & $\gamma$ \\
  &$200/500$& $100$ &$\sqrt{8/5}$&$1250$&$0.3$ \\ \hline
  $\tilde{h}^{(4)}$ in Fig.~\ref{fig:circuits_numerical}(c) [Eq.~(\ref{eq:h4t_num})] &$C_q$ [fF]& $L_q$ [pH] &$C_c$ [fF] & & \\
  &$500$& $100$ &$2$& &
\end{tabular}
\end{ruledtabular}
\end{table*}

\subsection{$g^{(4)}$ with single-junction coupler}
To calculate $g^{(4)}$, we use the circuit shown in Fig.~\ref{fig:original}.
The coupler has a single Josephson junction with a critical current $I_{0g}$,
and each KPO incorporates a SQUID.
According to Frattini et al.~\cite{Frattini:2018ahg}, the analytic formula for the nonlinearity is expressed as
\begin{equation}
  \hbar K_g = \frac{L_{Jg}^3}{\left( L_g +L_{Jg} \right)^3} \frac{e^2}{2C_g},
\end{equation}
where $e$ is the elementary charge,
and $L_{Jg}$ is the Josephson-junction inductance defined by $\varphi_0/I_{0g}$.
Here, $\varphi_0$ is the reduced flux quantum $\hbar/2e$.
The resonance frequency of the coupler is given by
\begin{equation}
  \omega_g = \frac{1}{\sqrt{C_g (L_g +L_{Jg})}}.
\end{equation}

For the parameters of the coupler, we adopt two parameter sets:
a KPO-like parameter set and transmon-like parameter set, as listed in Table~\ref{tab:parameters}.
The capacitance of the coupler mode, denoted as $\tilde{C}_g$ in Appendix A,
slightly differs from the cavity capacitance $C_g$.
However, since the difference is small, we use $C_g$ for the coupler capacitance.
Hereafter, we set the coupler frequency $\omega_g/2\pi = 10$ GHz.
At this frequency, we obtain $K_g/2\pi = 20.0$ MHz and $I_{0g} = 809$ nA with the KPO-like parameter set, and $K_g/2\pi = 172$ MHz and $I_{0g} = 135$ nA with the transmon-like parameter set, respectively.
As mentioned earlier, since $g^{(4)}$ is proportional to $K_g$,
we obtain a larger $g^{(4)}$ with the transmon-like parameter set.
Typically, as reflected in these parameter sets,
a critical current much smaller than $1$ $\mathrm{\mu}$A is required
to make $K_g$ large relative to a typical KPO nonlinearity
(e.g., $K_g/2\pi > 100$ MHz),
while the critical currents of Josephson junctions used in KPOs
are generally around $1$ $\mathrm{\mu}$A~\cite{Yamaji:2020pfg}.
Therefore, when using single junction couplers with large $g^{(4)}$s,
it is necessary to fabricate Josephson junctions of couplers with critical currents
that differ significantly from those of KPOs,
which leads to extra complexity in fabrication.
We can avoid this difficulty by using a frequency-tunable element such as a SQUID or quarton for the coupler
because a Kerr nonlinearity exceeding $100$ MHz and critical current value around $1$ $\mathrm{\mu}$A
can coexist at a resonance frequency used for a superconducting qubit (typically $5$--$10$ GHz).
However, these elements require individual flux tuning in addition to what is needed for KPOs,
which is undesirable for the operation of a large-scale device.

For the parameters of the KPOs, we use the same ones as those in the KPO-like parameter set, i.e., $C_q = 500$ fF and $L_q = 100$ pH.
We also use $C_q$ instead of $\tilde{C}_q$ in Appendix A for the KPO capacitance.

We assume that $\omega_j$s and $\omega_g$ satisfy the following relations:
\begin{equation}
  \omega_1=\omega_g +2\varepsilon,\qquad \left| \Delta_{1}\right| = 2\varepsilon,
\end{equation}
\begin{equation}
  \omega_2 = \omega_g -2 \varepsilon, \qquad \left| \Delta_{2} \right| = 2\varepsilon,
\end{equation}
\begin{equation}
  \omega_3 = \omega_g +\varepsilon, \qquad \left| \Delta_{3} \right| = \varepsilon,
\end{equation}
\begin{equation}
  \omega_4 = \omega_g -\varepsilon, \qquad \left| \Delta_{4}\right| = \varepsilon,
\end{equation}
where $\Delta_{j} = \omega_j -\omega_g$, and $\varepsilon$
is a parameter representing the detuning, which we call the unit detuning.
The $\omega_j$s also satisfy the condition in Eq.~(\ref{eq:freq_condition}).
Using the above conditions, we can rewrite the formula for $g^{(4)}$ as
\begin{equation}
  g^{(4)}=\frac{g_g^4}{2\varepsilon^4}K_g\label{eq:g4_num},
\end{equation}
\begin{equation}
  g_{j} \simeq \frac{1}{4} \frac{C_c}{\sqrt{C_q C_g}} \sqrt{\omega_j \omega_g} \simeq \frac{1}{4} \frac{C_c}{\sqrt{C_q C_g}} \omega_g \equiv g_g,
\end{equation}
where $g_{j}$ is the two-body coupling between KPO $j$ and the coupler.
We use $C_c = 1$ fF for the KPO-like coupler and $C_c = 1/\sqrt{5}$ fF for the transmon-like coupler
to adjust $g_g$s to be identical: $g_g/2\pi = 5$ MHz.

\subsection{$h^{(4)}$ with SQUID KPOs} \label{subsec:h4_SQUID}
\begin{figure*}[]
\includegraphics[width=18cm]{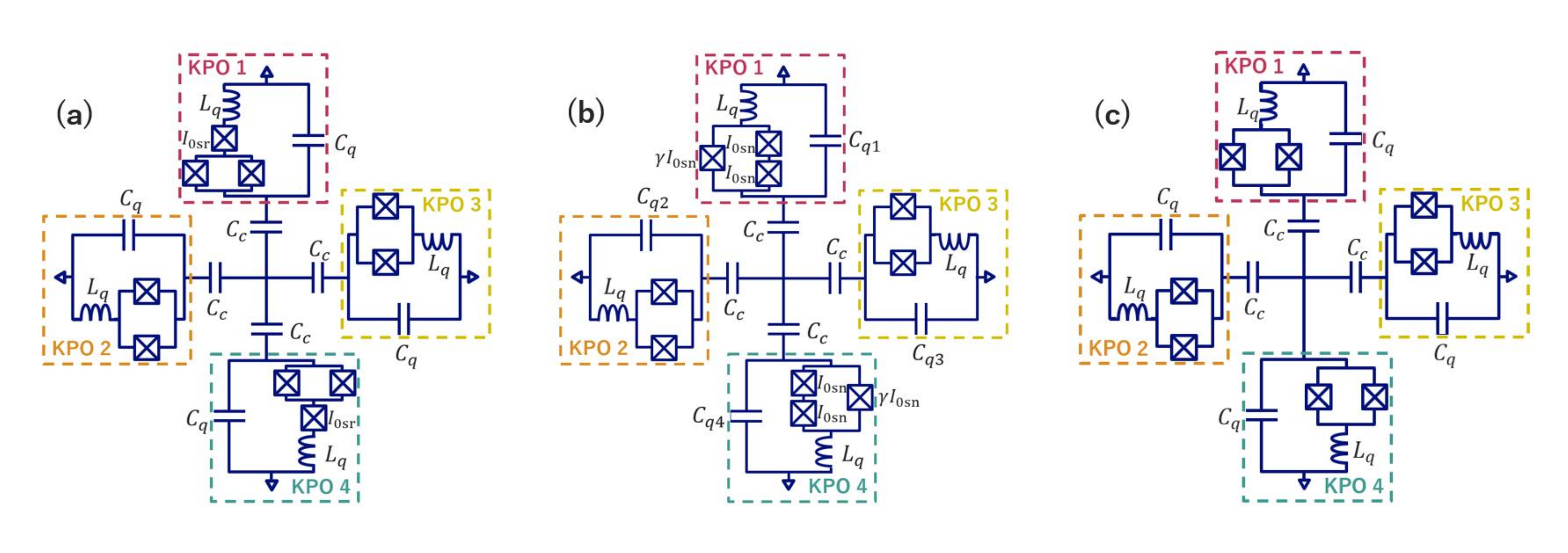}
\caption{\label{fig:circuits_numerical} Unit circuits for which the coupling constants are calculated in Sec.~\ref{sec:numerical}.
In (a), KPOs 1 and 4 each have a SQUID and single junction in series,
while KPOs 2 and 3 each have only a SQUID.
In (b), KPOs 1 and 4 each have a SNAIL, while KPOs 2 and 3 each have a SQUID.
In (c), all KPOs have a SQUID. Compared with the circuits in (a) and (b), the coupling capacitance for KPO 4 is eliminated.
}
\end{figure*}

Next, we consider $h^{(4)}$, which originates from KPO nonlinearities.
In this subsection, we assume the circuit shown in Fig.~\ref{fig:circuits_numerical}(a)
and parameter values listed in Table~\ref{tab:parameters}.
KPOs 1 and 4 each have a single junction with a critical current $I_{0\mathrm{sr}}$ in series with a SQUID, while KPOs 2 and 3 do not.
This configuration is necessary
because different values for the nonlinearities are required to make $h^{(4)}$ nonzero, as shown in Eq.~(\ref{eq:h4}).
The analytic formula for the Kerr nonlinearity of all KPOs is given as~\cite{Frattini:2018ahg}
\begin{equation}
  \hbar K_{q} = \frac{(L_{J\mathrm{sq}}^3 +L_{J\mathrm{sr}}^3)}{(L_q +L_{J\mathrm{sq}} +L_{J\mathrm{sr}})^3} \frac{e^2}{2C_q}, \label{eq:Kq}
\end{equation}
where $L_{J\mathrm{sq}}$ and $L_{J\mathrm{sr}}$ ($=\varphi_0/I_{0\mathrm{sr}}$) are the Josephson inductances of the SQUID and single junction, respectively.
The $L_{J\mathrm{sq}}$ at $10$ GHz is determined to be $L_{J\mathrm{sq}}=187$ pH for KPOs 1 and 4,
and $L_{J\mathrm{sq}}=407$ pH for KPOs 2 and 3.
The Kerr nonlinearities $K_j/2\pi$ at the frequency
can be obtained as $K_1/2\pi = K_4/2\pi = 5.1$ MHz and $K_2/2\pi = K_3/2\pi = 20$ MHz
by using Eq.~(\ref{eq:Kq}).
To satisfy the condition in Eq.~(\ref{eq:freq_condition}), the following relations are assumed for the resonance frequencies:
\begin{equation}
  \omega_2=\omega_1-3\varepsilon, \qquad \left|\mathrm{\Delta}_{12}\right|=3\varepsilon, \label{eq:freq_diff_kerr2}
\end{equation}
\begin{equation}
  \omega_3=\omega_1-\varepsilon, \qquad \left|\mathrm{\Delta}_{13}\right|=\varepsilon, \label{eq:freq_diff_kerr3}
\end{equation}
\begin{equation}
  \omega_4=\omega_1-2\varepsilon, \qquad \left|\mathrm{\Delta}_{14}\right|=2\varepsilon, \label{eq:freq_diff_kerr4}
\end{equation}
where we set $\omega_1/2\pi = 10$ GHz.
The coupling in Eq.~(\ref{eq:h4}) is rewritten as
\begin{equation}
  h^{(4)} = h_q^3 \frac{\left( K_2 -K_1 \right) + 3 \left( K_3 -K_4 \right)}{3 \varepsilon^3}, \label{eq:approx_h4}
\end{equation}
where $h_q$ is the approximated coupling constant between the KPOs:
\begin{equation}
  h_{jk} \simeq \frac{C_c}{8 C_q} \sqrt{\omega_j \omega_k} \simeq \frac{C_c}{8 C_q} \omega_1 \equiv h_q.
\end{equation}
We use $C_c = 2$ fF to make $h_q$ identical to $g_g$: $h_q/2\pi = g_g/2\pi = 5$ MHz.
It is worth noting that with the circuit in Fig.~\ref{fig:circuits_numerical}(a),
we do not need to worry about the coupler mode.
This is a significant advantage over the circuit in Fig.~\ref{fig:original}, which requires $\braket{a_g} = \braket{a_g^\dagger a_g} = 0$~\cite{Puri:2017hqg}.

\subsection{$h^{(4)}$ with combination of SQUID KPOs and SNAIL KPOs}
As mentioned in Sec.~\ref{sec:four},
we can achieve a larger coupling magunitude for $h^{(4)}$
by using a junction element which has a Kerr nonlinearity
with an opposite sign compared with a SQUID, such as a SNAIL.
Charge-driven SNAIL KPOs (KPOs with SNAILs) in such a parameter regime have been investigated~\cite{Venkatraman:2022dxu}.
We now calculate $h^{(4)}$ when using SNAILs in combination with SQUIDs.
The circuit considered in this subsection is shown in Fig.~\ref{fig:circuits_numerical}(b).
KPOs 1 and 4 each have a SNAIL, and KPOs 2 and 3 each have a SQUID.
In the SNAIL, two junctions with a critical current $I_{0\mathrm{sn}}$ are in parallel with one junction with $\gamma I_{0\mathrm{sn}}$,
where $\gamma$ is a parameter satisfying $0 < \gamma < 1$.
We adopt $\gamma=0.3$ to implement a system with feasible parameters, such as the critical current of the junction around 1 $\mathrm{\mu}$A.

In order to calculate $h^{(4)}$, we need to know $K_j$s, which each depend on $\omega_j$s,
but $K_1$ and $K_4$ cannot be expressed as analytical functions of $\omega_1$ and $\omega_4$, respectively.
From numerical calculations, we find that $K_1$ and $K_4$ are approximately proportional to $\omega_1$ and $\omega_4$,
correspondingly, when $\omega_1/2\pi$ and $\omega_4/2\pi$ are just below 10 GHz.
We then fit $K_1$ and $K_4$ as linear functions of $\omega_1$ and $\omega_4$,
respectively, and obtain one-to-one correspondences between them.
Details of the calculation are provided in Appendix C.

Since the capacitances of the KPOs are not identical, we cannot use the approximated formula in Eq.~(\ref{eq:approx_h4}).
Instead, we use $h^{(4)}$ in Eq.~(\ref{eq:original-h4})
and assume the following relations:
\begin{equation}
  h_{12} = h_{13} = h_{24} = h_{34} \equiv h_{\mathrm{\scalebox{0.7}{QN}}},
\end{equation}
\begin{equation}
  h_{14} \equiv h_{\mathrm{\scalebox{0.7}{NN}}},
\end{equation}
\begin{equation}
  h_{23} \equiv h_{\mathrm{\scalebox{0.7}{QQ}}},
\end{equation}
where the subscript $\mathrm{Q}$ represents a SQUID KPO (a KPO with a SQUID), and $\mathrm{N}$ represents a SNAIL KPO.
Namely, $h_{\mathrm{\scalebox{0.7}{QN}}}$, for example, denotes the coupling constant between a SQUID KPO and SNAIL KPO.
We use $C_c = \sqrt{8/5}$ fF to make $h_{\mathrm{\scalebox{0.7}{QN}}}$ identical to $g_g$: $h_{\mathrm{\scalebox{0.7}{QN}}}/2\pi = g_g/2\pi = 5$ MHz.
The $h^{(4)}$ is rewritten as
\begin{equation}
  h^{(4)} = h_{\mathrm{\scalebox{0.7}{QN}}}^2 \frac{-h_{\mathrm{\scalebox{0.7}{NN}}} \left( K_1 +3K_4\right) +h_{\mathrm{\scalebox{0.7}{QQ}}}\left( K_2 +3K_3\right)}{3\varepsilon^3}\label{eq:h4_SNAIL},
\end{equation}
where the relations among the resonance frequencies
presented in Eqs.~(\ref{eq:freq_diff_kerr2})--(\ref{eq:freq_diff_kerr4})
are adopted.
We see that by preparing negative $K_1$ and $K_4$ with SNAIL KPOs and positive $K_2$ and $K_3$ with SQUID KPOs,
we can obtain a larger magnitude for $h^{(4)}$ compared with the case using only SQUID KPOs.
Since a SNAIL can have both positive and negative signs for its Kerr nonlinearity,
we can also construct the unit circuit with only SNAILs.

\subsection{$\tilde{h}^{(4)}$ with SQUID KPOs}
We consider the circuit in Fig.~\ref{fig:circuits_numerical}(c) and $\tilde{h}^{(4)}$ on the basis of the formula in Eq.~(\ref{eq:one_origination}).
In this circuit, direct couplings between KPOs 1, 2, and 3 are suppressed,
thus only $K_4$ contributes to the formula for the four-body interaction.
We use the same KPO parameters as those in Subsec~\ref{subsec:h4_SQUID}.
Assuming the relations of $\omega_j$s in Eqs.~(\ref{eq:freq_diff_kerr2})--(\ref{eq:freq_diff_kerr4}),
we obtain the following formula for $\tilde{h}^{(4)}$:
\begin{equation}
  \tilde{h}^{(4)} = -h_q^3 \frac{K_4}{\varepsilon^3}\label{eq:h4t_num},
\end{equation}
where $C_c=2$ fF is used to achieve $h_q/2\pi = g_g/2\pi = 5$ MHz.

\subsection{Numerical results}
\begin{figure}[tb]
\begin{center}
\includegraphics[width=8cm]{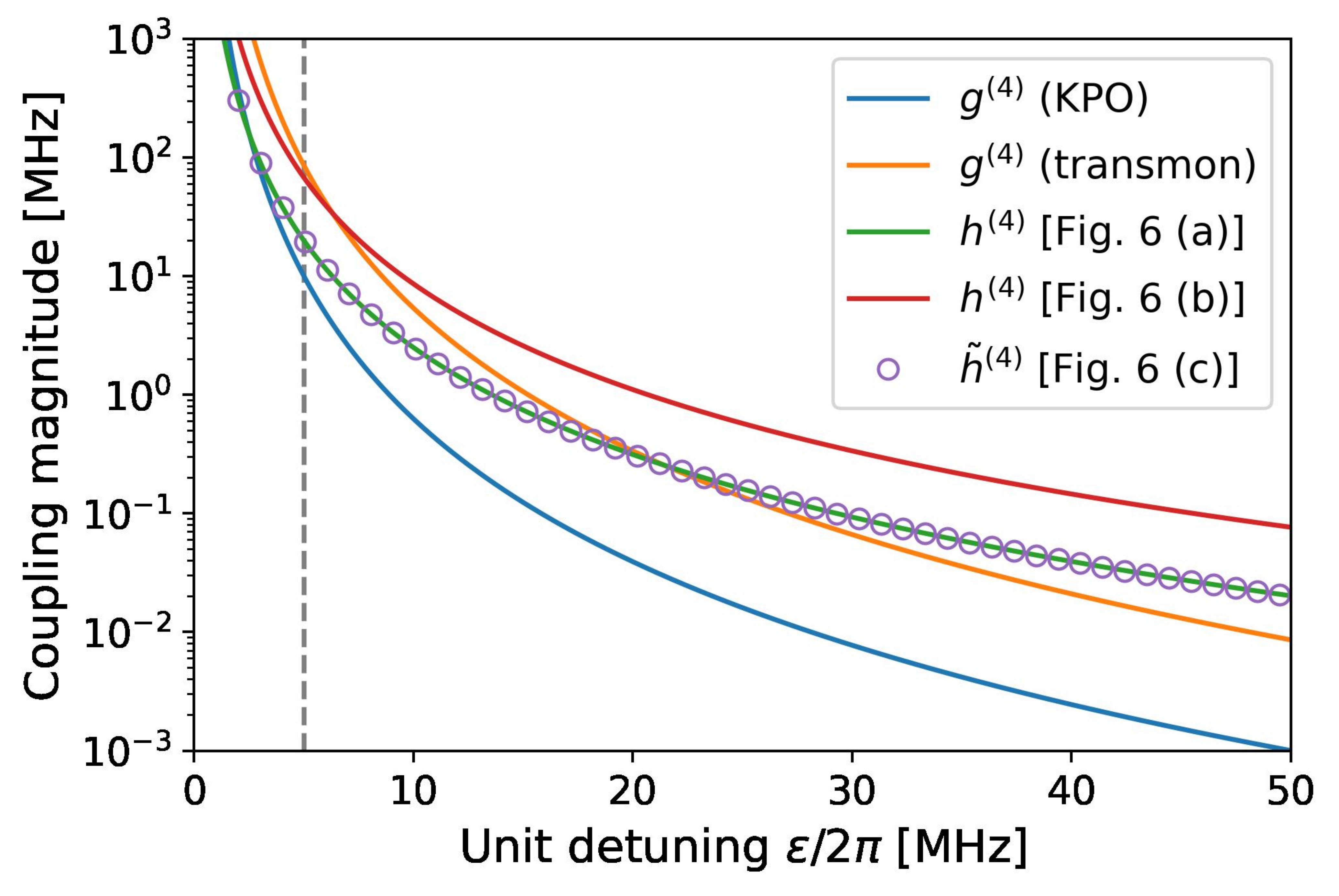}
\caption{
  Four-body couplings as a function of unit detuning $\varepsilon$.
  The blue curve shows the magnitude for $g^{(4)}/2\pi$ with the KPO-like coupler,
  and the orange curve shows that with the transmon-like coupler.
  The green and red curves show the magnitudes for $h^{(4)}/2\pi$ with
  the circuits shown in Fig.~\ref{fig:circuits_numerical}(a)
  and (c), respectively.
  The purple open circles show the magnitudes for $\tilde{h}^{(4)}/2\pi$ with the circuit shown in Fig.~\ref{fig:circuits_numerical}(c),
  and it almost overlaps with the green curve.
  The gray vertical line represents the detuning which is equal to $g_g/2\pi = h_q/2\pi = h_{\mathrm{\scalebox{0.7}{QN}}}/2\pi = 5$ MHz.
}
\label{fig:all_fbc_log}
\end{center}
\end{figure}

We present the numerical calculations of the four-body couplings.
Figure~\ref{fig:all_fbc_log} shows the values for $g^{(4)}$, $h^{(4)}$, and $\tilde{h}^{(4)}$ with the five parameter sets in Table~\ref{tab:parameters}.
We first focus on $g^{(4)}$s.
Since the Kerr nonlinearity of the transmon-like coupler is larger than that of the KPO-like coupler,
the magnitude for $g^{(4)}$ with the transmon-like coupler (orange curve) is larger than that of the KPO-like coupler (blue curve).
We need to make the unit detuning $\varepsilon$ small to obtain large magnitudes for $g^{(4)}$,
as it is proportional to $\varepsilon^{-4}$.
However, we have to keep $\varepsilon$ large compared to $g_g$, $h_q$, and $h_{\mathrm{\scalebox{0.7}{QN}}}$
(vertical dashed line in Fig.~\ref{fig:all_fbc_log}),
otherwise the approximation used to derive the coupling magnitudes become inaccurate.

Next, we focus on $h^{(4)}$ and $\tilde{h}^{(4)}$.
Similar to the results of $g^{(4)}$s, small $\varepsilon$ is required to obtain a large magnitude for $h^{(4)}$.
It can be seen that larger magnitudes are obtained compared with $g^{(4)}$ when $\varepsilon$ is relatively large ($\varepsilon/2\pi > 25$ MHz)
since $h^{(4)}$ and $\tilde{h}^{(4)}$ are at the third order of the perturbation, while $g^{(4)}$ is at the fourth order.
Using the combination of SQUIDs and SNAILs, the coupling magnitude (red curve) is larger than that of $h^{(4)}$ with only SQUID KPOs (green curve).
We find that the coupling magnitude for $\tilde{h}^{(4)}$ (purple open circles) is nearly identical to that for $h^{(4)}$ with only SQUIDs
for this parameter set.
This result indicates that, by modifying the capacitance geometry,
a relatively large four-body coupling can be achieved
without introducing differences in KPO nonlinearities.

\section{Experimental demonstration}\label{sec:exp}

We experimentally investigated four-body interactions
by measuring the correlation between the oscillation states of KPOs.
We followed the experimental method introduced by our previous work~\cite{Yamaji:2022ikh},
where the two-body correlation between capacitively coupled KPOs was investigated.

\subsection{Device}
\begin{figure}[tb]
\begin{center}
\includegraphics[width=8cm]{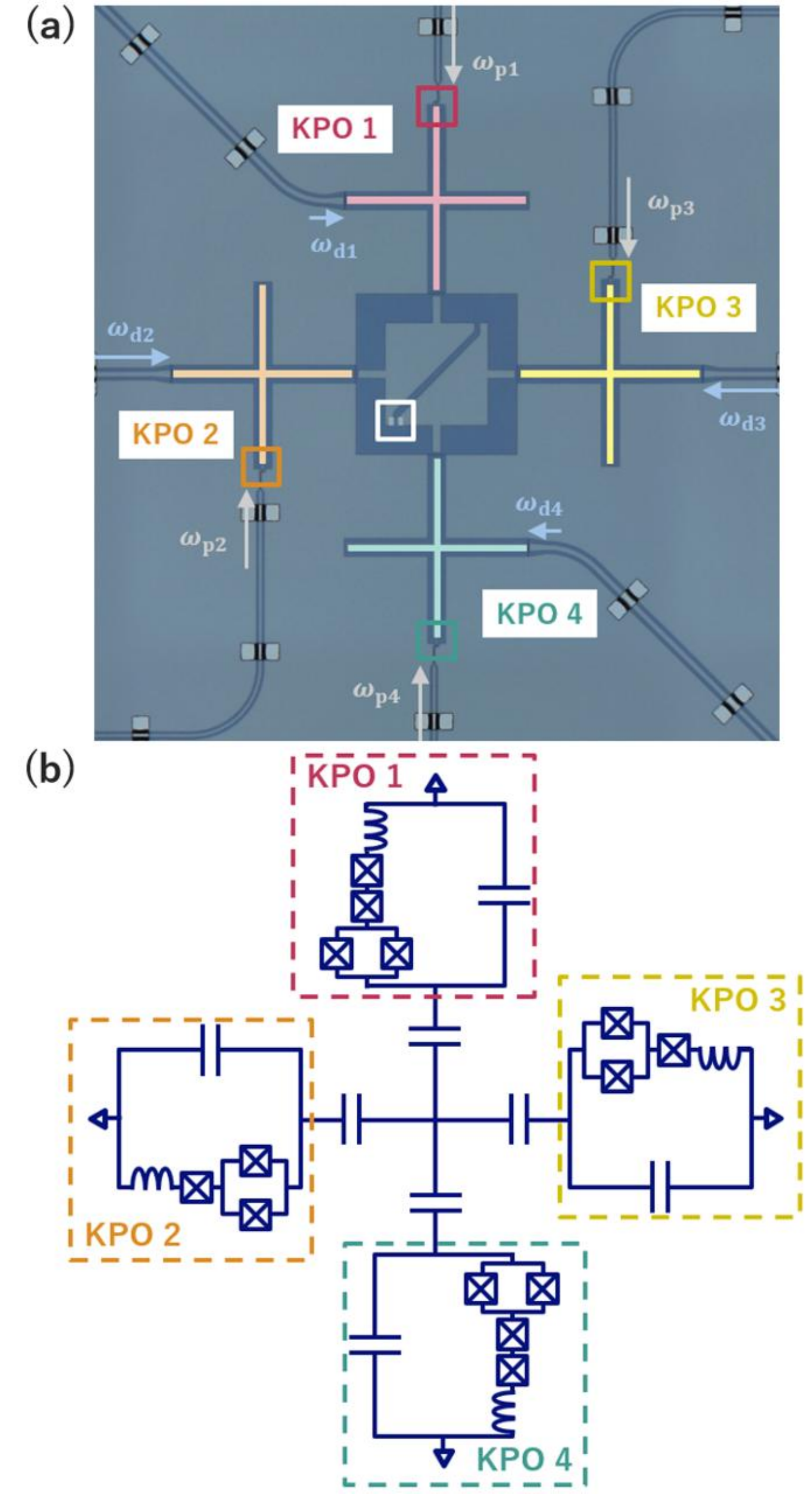}
\caption{
(a)  
Optical microscope image of the device, including four KPOs and a coupler.
We label the KPOs as KPO 1 to KPO 4.
The coupler consists of two capacitor pads
connected by a Josephson junction (white box).
Each KPO is connected to two lines:
an I/O line for coherent drive at drive frequency $\omega_{\mathrm{d}j}$
and a pump line for parametric pumping
at pump frequency $\omega_{\mathrm{p}j}$ and dc flux bias.
A SQUID (colored box) is inductively coupled to each pump line.
Both I/O and pump lines are equipped with airbridges to suppress ac/dc crosstalk and unwanted slotline modes. 
(b) Circuit model of the device.
The Josephson junction in the coupler is regarded as a short
because the resonance frequency of the coupler is much higher than those of KPOs.
}\label{fig:chip_and_model}
\end{center}
\end{figure}

Figure~\ref{fig:chip_and_model}(a) shows an optical microscope image of the device.
There are four KPOs around the coupler located at the center.
While the resonance frequencies of the KPOs are around 9.3 GHz at the operating point,
the coupler's resonance frequency is around 20 GHz.
Therefore, the four-body coupling originating from the coupler nonlinearity is negligible
since the coupler's detuning from the KPO frequencies is much larger than any coupling magnitudes in the circuit.
Each KPO is connected to a pump line and input/output (I/O) line.
The pump lines are inductively coupled to each KPO's SQUID.
Each KPO oscillates when it is driven
at a pump frequency $\omega_{\mathrm{p}j} \approx 2\omega_j$ through this line.
The resonance frequency of each KPO can be controlled by applying dc flux to its SQUID through the same line.
The I/O lines are capacitively coupled to each KPO.
We coherently drive each KPO with a drive frequency $\omega_{\mathrm{d}j} = \omega_{\mathrm{p}j}/2$
using this line.
The I/O lines are also used for readout of the output signals.
See Appendix E for details of the fabrication.

Figure~\ref{fig:chip_and_model}(b) shows the circuit diagram of the device, where the degree of freedom of the coupler is omitted.
KPOs 1 and 4 each have a SQUID and two series junctions, while KPOs 2 and 3 each have a SQUID and one series junction.
This design introduces differences in the KPO nonlinearities,
as described in the previous section.
Thus, a nonzero four-body coupling is expected on the basis of Eq.~(\ref{eq:approx_h4}).
The set resonance frequencies and measured Kerr nonlinearities are listed in Table~\ref{tab:kpo_para},
latter of which are estimated from the measurement of reflection coefficient with
varied probe powers~\cite{Yamaji:2020pfg}.
\begin{table*}
\caption{\label{tab:currents}
Measured parameters of the resonance frequencies, Kerr nonlinearities, oscillation amplitudes,
and magnitudes of coherent drives.
}
\label{tab:kpo_para}
\begin{ruledtabular}
\begin{tabular}{ccccccc}
   & $\tilde{\omega}_j/2\pi$ [GHz] &$\tilde{K}_j/2\pi$ [MHz]&$\alpha_j$ &$\epsilon_j/2\pi$ [MHz] \\
  KPO 1 &$9.33$&$10.4$ & $5.9$ & $1.3\times10^2$ \\
  KPO 2 &$9.31$&$15.2$&$4.5$&$5.3$ \\
  KPO 3 &$9.35$&$13.2$&$1.3$&$29$ \\
  KPO 4 &$9.29$&$10.0$&$5.3$& $4.5$ \\
\end{tabular}
\end{ruledtabular}
\end{table*}

The two-body couplings $h_{jk}$ can be estimated from the avoided crossings between the KPOs,
which were measured using another device with identical geometry.
The measured values are listed in Table~\ref{tab:two-body}.
\begin{table*}
\caption{Measured magnitudes of $h_{jk}$s estimated from avoided crossings between the KPOs.
These values were evaluated using another device with identical geometry.}
\label{tab:two-body}
\begin{ruledtabular}
\begin{tabular}{ccccccc}
  $h_{12}/2\pi$ [MHz] & $h_{13}/2\pi$ [MHz]& $h_{14}/2\pi$ [MHz] & $h_{23}/2\pi$ [MHz] & $h_{24}/2\pi$ [MHz] & $h_{34}/2\pi$ [MHz] & \\
  $5.8$ & $4.4$ & $2.5$ & $4.5$ & $4.4$ & $4.7$ & \\ 
\end{tabular}
\end{ruledtabular}
\end{table*}
Using these values, we roughly estimated the four-body coupling as $|h^{(4)}|/2\pi = 0.1$ MHz on the basis of Eq.~(\ref{eq:original-h4}).
Note that we used the measured values of the Kerr nonlinearities $\tilde{K}_j/2\pi$
and resonance frequencies $\tilde{\omega}_j/2\pi$ listed in Table~\ref{tab:kpo_para}
as $K_j/2\pi$ and $\omega_j/2\pi -K_j/2\pi$, respectively.
They are distinct when the KPO frequencies are close to each other,
as in the present case
($\varepsilon/2\pi \approx 20$ MHz).
See Appendix~B for the perturbative expressions for
$\tilde{\omega}_j$ and $\tilde{K}_j$.

\subsection{Time-domain measurement of KPO}
\begin{figure}[tb]
\begin{center}
\includegraphics[width=9cm]{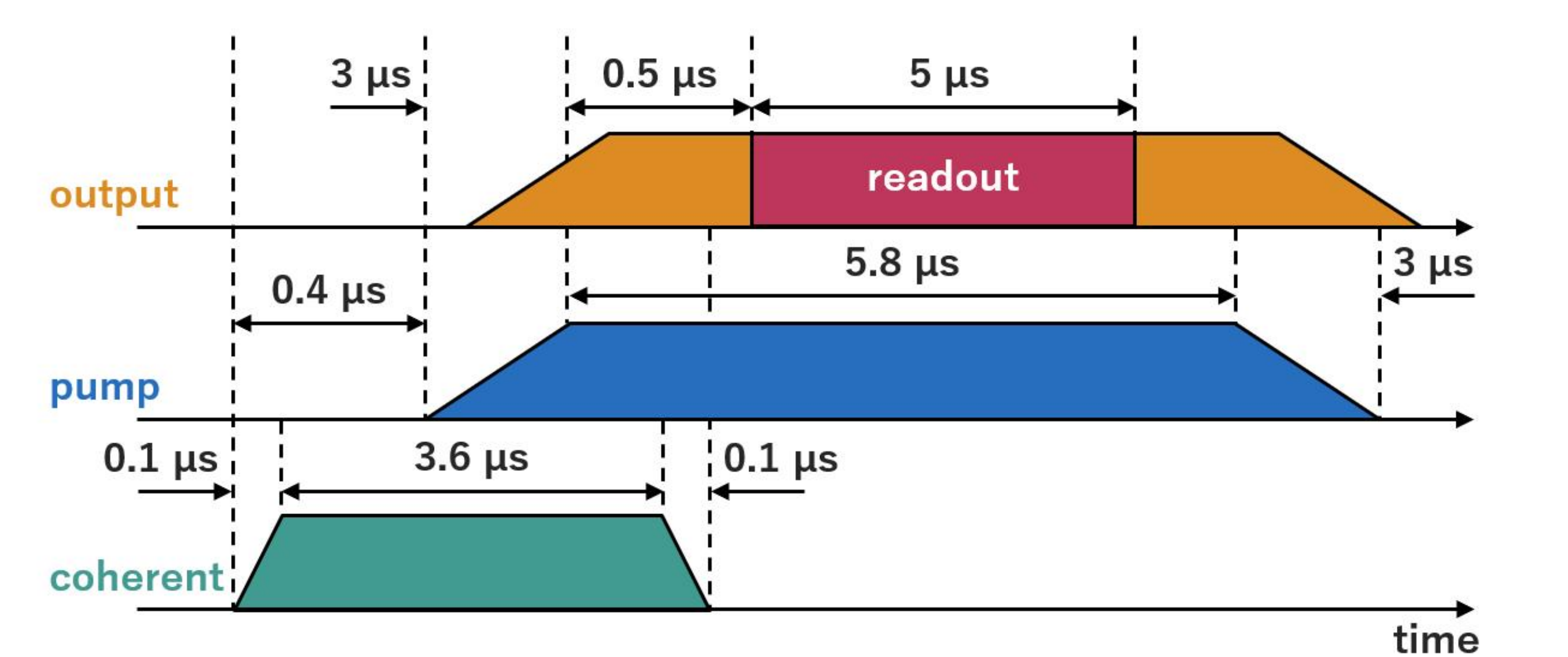}
\caption{
Pulse sequence of output, pump drive, and coherent drive.
}\label{fig:pulse_sequence}
\end{center}
\end{figure}

In the following, we conducted time-domain measurement by applying pulsed pump and coherent drives for the KPOs.
Figure~\ref{fig:pulse_sequence} shows the pulse sequence of the measurement.
The output signal, which represents the oscillation state of a KPO,
was recorded for a readout time of 5 $\mathrm{\mu}$s.
When Fourier transform is applied to the measurement data,
two peaks are observed in the in-phase and quadrature (IQ) plane.
The histogram of the plots has two peaks
corresponding to the two coherent states of the parametric oscillation
which have equal magnitude and phases shifted by $\pi$.
When the states are degenerate, their occurrence probabilities are approximately the same.
These probabilities can be biased by applying a coherent drive~\cite{Lin:2014wsg}.
When applying a coherent drive, as we do in Sec.~\ref{subsec:state-control},
we read the output signal after turning off the drive
because otherwise we cannot discriminate the output signal from the reflected coherent drive. 
The measurement setup was the same as that used in our previous paper~\cite{Yamaji:2022ikh},
where the details can be found.
We summarize the room temperature electronics in Appendix E.

\subsection{Hamiltonian and effective energies}\label{subsec:hamiltonian}
In the rotating frame at half the pump frequencies [see Eq.~(\ref{eq:rf-trans})], the Hamiltonian of the four-KPO system is written as
\begin{equation}
  H = H_{\mathrm{KPO}} +H_{\mathrm{four}} +H_{\mathrm{res}1} +H_{\mathrm{res}2} +H_{\mathrm{coh}},
\end{equation}
\begin{equation}
  H_{\mathrm{KPO}}/\hbar = \sum_j^4 \left\{ \tilde{\delta_j} a_j^\dagger a_j -\frac{\tilde{K}_j}{2} a_j^{\dagger2} a_j^2 +\frac{p_j}{2} \left( a_j^2 +a_j^{\dagger2}\right)\right\}\label{eq:H_KPO},
\end{equation}
\begin{equation}
  H_{\mathrm{four}}/\hbar = -h^{(4)} e^{i\theta_{\mathrm{p}}/2} a_1^\dagger a_2^\dagger a_3 a_4 +\mathrm{h.c.},
\end{equation}
\begin{equation}
  H_{\mathrm{res}1}/\hbar = g^\prime a_1^\dagger a_4^\dagger a_2^2 +g^{\prime\prime} a_1^2 a_2^\dagger a_3^\dagger +\mathrm{h.c.} \label{eq:H_res1},
\end{equation}
\begin{equation}
  H_{\mathrm{res}2}/\hbar = g^{\prime\prime\prime} a_1^{\dagger3} a_3^2 a_4 +g^{\prime\prime\prime\prime} a_2^{\dagger3} a_3 a_4^2 +\mathrm{h.c.} \label{eq:H_res2},
\end{equation}
\begin{equation}
  H_{\mathrm{coh}}/\hbar = -i \sum_{j=1}^4 \epsilon_j \left( a_j e^{i\theta_{\mathrm{d}j}} -a_j^\dagger e^{-i \theta_{\mathrm{d}j}} \right) \label{eq:coherent},
\end{equation}
where $H_{\mathrm{KPO}}$ is the Hamiltonian of the four KPOs,
$H_{\mathrm{four}}$ is that of the four-body interaction,
$H_{\mathrm{res}1}$ and $H_{\mathrm{res}2}$ are those of residual interactions,
which will be discussed later,
and $H_{\mathrm{coh}}$ is that of the coherent drive
with $\omega_{\mathrm{d}j} = \omega_{\mathrm{p}j}/2$.
We assume the four-body mixing condition $\omega_{\mathrm{p}1} +\omega_{\mathrm{p}2} = \omega_{\mathrm{p}3} +\omega_{\mathrm{p}4}$
and omit the rotating terms.

In $H_{\mathrm{KPO}}$, $\tilde{\delta}_j$ represents the frequency detuning in the rotating frame
$\tilde{\delta}_j = \tilde{\omega}_j -\omega_{\mathrm{p}j}/2$ for KPO $j$.
We set $\tilde{\delta}_j/2\pi = 60$ MHz for all $j$ in the experiment.
The third terms in Eq.~(\ref{eq:H_KPO}) correspond to the pump drives.
The $g^\prime$, $g^{\prime \prime}$, $g^{\prime \prime \prime}$, and $g^{\prime \prime \prime \prime}$
in $H_{\mathrm{res}1}$ and $H_{\mathrm{res}2}$ are the coupling constants of the residual interactions.
The $\epsilon_j$ and $\theta_{\mathrm{d}j}$ in $H_{\mathrm{coh}}$ are the amplitude and phase of
the coherent drive for KPO $j$.
The amplitude can be expressed as $\epsilon_j = \sqrt{P_{\mathrm{d}j} \kappa_{\mathrm{e}j}/\hbar \omega_{\mathrm{d}j}}$,
where $P_{\mathrm{d}j}$ is the power of the coherent drive at the device and
$\kappa_{\mathrm{e}j}$ is the external decay rate to the I/O line~\cite{Yamaji:2020pfg}.
The values of $\epsilon_j$s used in the experiment described
in the following subsection are listed in Table~\ref{tab:kpo_para}.

The oscillation state of each KPO can be approximated as a coherent state $\ket{\pm \alpha_j}$
with a oscillation amplitude $\alpha_j$.
The entire system is represented by a product state
$\ket{s_1 \alpha_1} \ket{s_2 \alpha_2} \ket{s_3 \alpha_3} \ket{s_4 \alpha_4}$,
where $s_j = \pm 1$ represents an Ising spin~\cite{Yamaji:2022ikh}.

The eigenenergies corresponding to the interaction Hamiltonians $H_{\mathrm{four}}$,
$H_{\mathrm{res}1}$, and $H_{\mathrm{res}2}$,
and the coherent drive Hamiltonian $H_{\mathrm{coh}}$ are respectively expressed as
\begin{equation}
  E_{\mathrm{four}} = \breve{h}^{(4)} s_1 s_2 s_3 s_4 \label{eq:E4},
\end{equation}
\begin{equation}
  E_{\mathrm{res}1} = \breve{g}^\prime s_1 s_4 +\breve{g}^{\prime \prime} s_2 s_3 \label{eq:E_res1},
\end{equation}
\begin{equation}
  E_{\mathrm{res}2} = \breve{g}^{\prime \prime \prime} s_1 s_4 +\breve{g}^{\prime \prime \prime \prime} s_2 s_3 \label{eq:E_res2},
\end{equation}
\begin{equation}
  E_{\mathrm{coh}} = \sum_{j=1}^4 \breve{\epsilon}_j s_j \label{eq:E1},
\end{equation}
where the coefficients are given by:
\begin{equation}
  \breve{h}^{(4)} = -2h^{(4)} \alpha_1 \alpha_2 \alpha_3 \alpha_4 \cos \frac{\theta_{\mathrm{p}}}{2} \label{eq:h4_b},
\end{equation}
\begin{equation}
  \breve{g}^{\prime} = 2g^\prime \alpha_1 \alpha_4 \alpha_2^2 \cos \left( \frac{\theta_{\mathrm{p}1}}{2} +\frac{\theta_{\mathrm{p}4}}{2} -\theta_{\mathrm{p}2}\right),
\end{equation}
\begin{equation}
  \breve{g}^{\prime \prime} = 2g^{\prime \prime} \alpha_1^2 \alpha_2 \alpha_3 \cos\left( \theta_{\mathrm{p}1} -\frac{\theta_{\mathrm{p}2}}{2} -\frac{\theta_{\mathrm{p}3}}{2} \right),
\end{equation}
\begin{equation}
  \breve{g}^{\prime \prime \prime} = 2g^{\prime \prime \prime} \alpha_1^3 \alpha_3^2 \alpha_4 \cos \left(\frac{3}{2}\theta_{\mathrm{p}1} -\theta_{\mathrm{p}3} -\frac{\theta_{\mathrm{p}4}}{2} \right) \label{eq:high-res-coeff1},
\end{equation}
\begin{equation}
  \breve{g}^{\prime \prime \prime \prime} = 2g^{\prime \prime \prime \prime} \alpha_2^3 \alpha_3 \alpha_4^2 \cos \left( \frac{3}{2} \theta_{\mathrm{p}2} -\frac{\theta_{\mathrm{p}3}}{2} -\theta_{\mathrm{p}4}\right) \label{eq:high-res-coeff2},
\end{equation}
\begin{equation}
  \breve{\epsilon}_j = 2\epsilon_j \alpha_j \sin \theta_{\mathrm{d}j} \label{eq:coh-coeff}.
\end{equation}
The $\theta_{\mathrm{p}j}$ is the phase of the pump drive for KPO $j$ and
we define $\theta_{\mathrm{p}} = \theta_{\mathrm{p}1} +\theta_{\mathrm{p}2} -\theta_{\mathrm{p}3} -\theta_{\mathrm{p}4}$.
For now, we consider $E_{\mathrm{four}}$ only.
Since $E_{\mathrm{four}}$ depends on the parity of the spin product $s_1 s_2 s_3 s_4$,
we experimentally investigated this parity to verify the presence of the four-body interaction.
The even-parity state is defined as the product state with $s_1 s_2 s_3 s_4 = +1$,
and the odd-parity state is defined as the product state with $s_1 s_2 s_3 s_4 = -1$.
We can control the occurrence probabilities of these parity states by varying the pump phases
on the basis of Eq.~(\ref{eq:h4_b}).

\subsection{Four-body correlation}\label{subsec:correlation}
\begin{figure*}[]
\includegraphics[width=14cm]{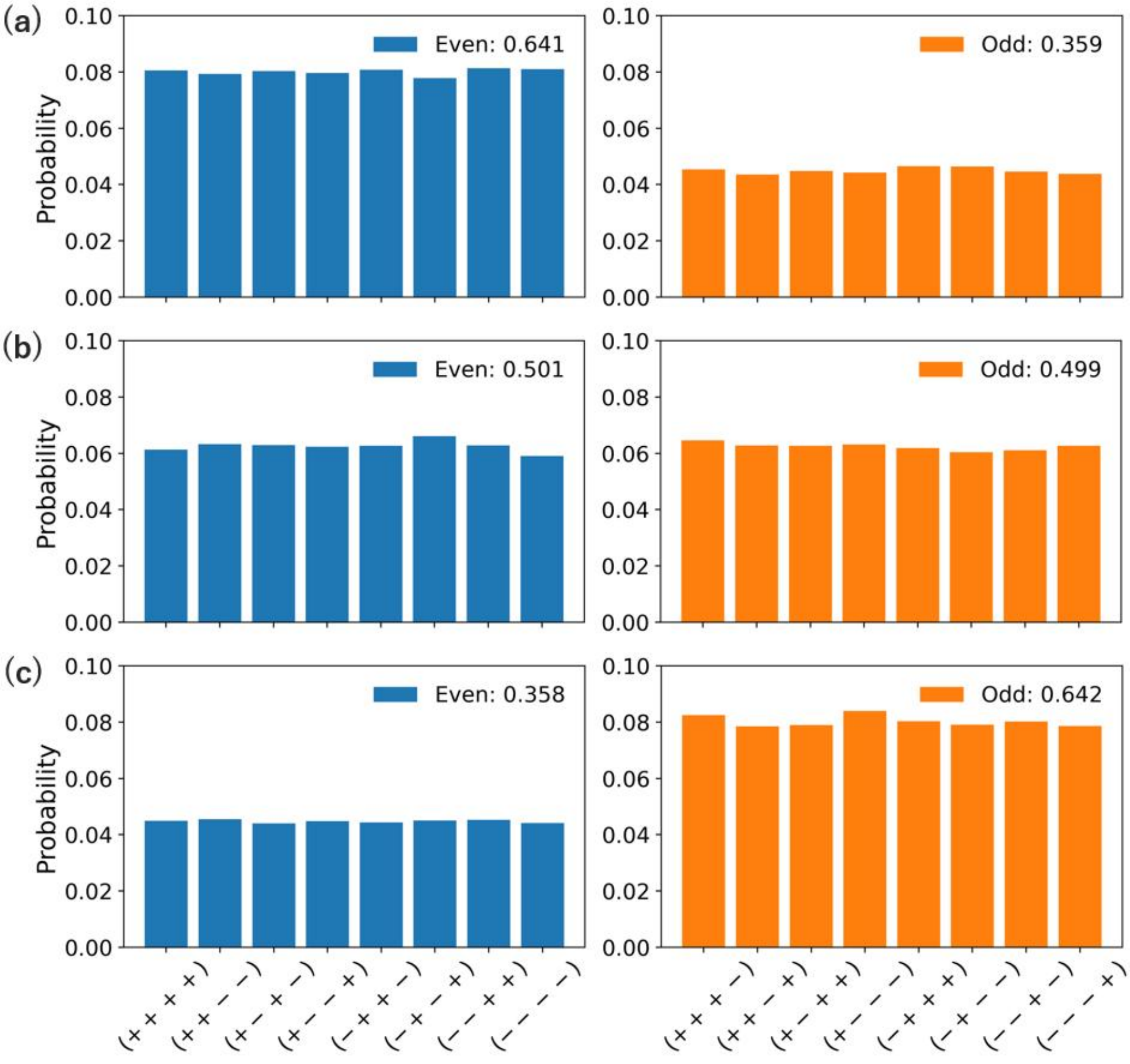}
\caption{\label{fig:probs_bar} 
Occurrence probabilities of four KPO states
with different $\theta_{\mathrm{p}}$ of (a) 0, (b) $\pi$, and (c) $2\pi$.
The probabilities of the even-parity states are represented in blue,
and those of the odd-parity states are in orange.
The $+$ ($-$) in the X-axis labels represents the $+1$ ($-1$) state.
For each $\theta_{\mathrm{p}}$ and each KPO, $4\times 10^4$ pump pulses were applied,
and the probabilities were calculated.
The statistical error is not shown since it is almost negligible ($< 1.5 \times 10^{-3}$).
}
\end{figure*}

Figure~\ref{fig:probs_bar} shows the experimental data of
occurrence probabilities of the four-KPO states,
where KPO 1, 2, 3, and 4 are parametrically driven with
$\omega_{\mathrm{p}1}/2\pi=2\times9.270$ GHz, $\omega_{\mathrm{p}2}/2\pi=2\times9.249$ GHz, $\omega_{\mathrm{p}3}/2\pi=2\times9.290$ GHz, and $\omega_{\mathrm{p}4}/2\pi=2\times9.229$ GHz, respectively.
These pump frequencies meet the four-body mixing condition ($\omega_{\mathrm{p1}} +\omega_{\mathrm{p2}} = \omega_{\mathrm{p}3} +\omega_{\mathrm{p}4}$).
We did not use coherent drives in this experiment.
There are $2^4=16$ states in total, divided into even-parity states (blue)
and odd-parity states (orange).
Figure~\ref{fig:probs_bar}(a) shows the probabilities with $\theta_\mathrm{p} = 0$
and $\breve{h}^{(4)} = -2h^{(4)} \alpha_1 \alpha_2 \alpha_3 \alpha_4 < 0$,
where the even-parity states are favored.
The figure indicates that the total probability of the even-parity states ($64.1\pm0.2$\%) is
higher than that of the odd-parity states ($35.9\pm0.2$\%).
This biasing can be controlled by varying $\theta_{\mathrm{p}}$.
Figure~\ref{fig:probs_bar}(b) shows the probabilities with $\theta_\mathrm{p} = \pi$,
where the biasing vanishes due to $\breve{h}^{(4)} = 0$.
The odd-parity states are favored with $\theta_\mathrm{p} = 2\pi$,
as shown in Fig.~\ref{fig:probs_bar}(c).

\begin{figure*}[]
\includegraphics[width=18cm]{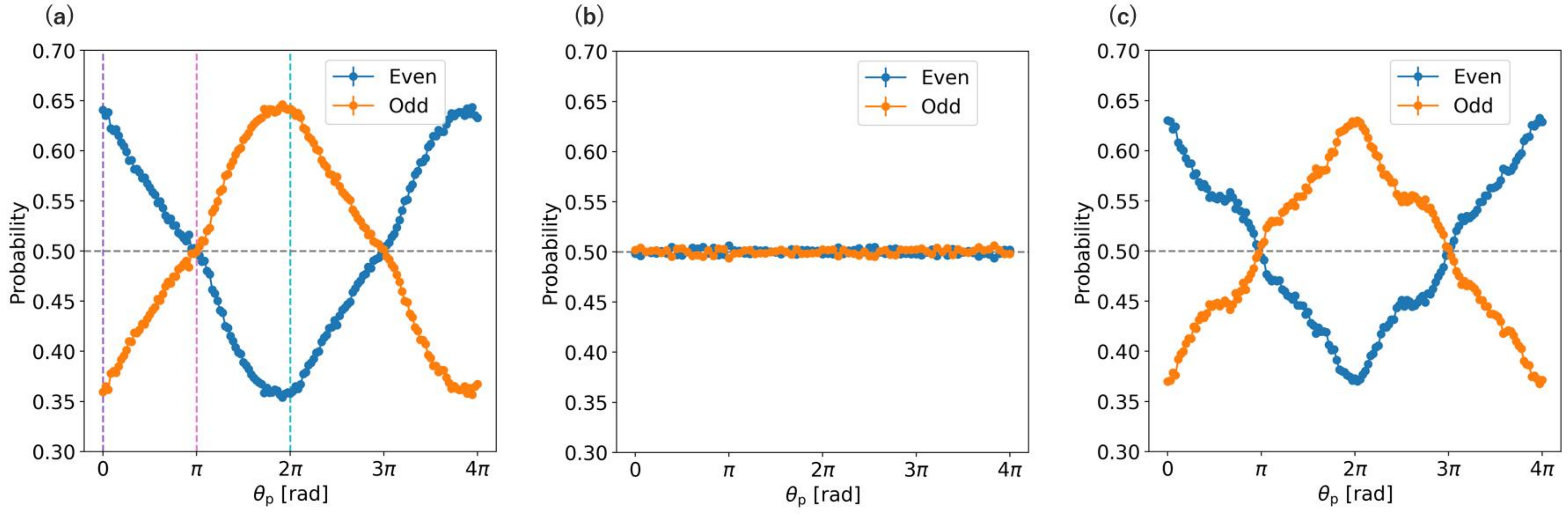}
\caption{\label{fig:even_odd_prob} 
Probability of parity as a function of $\theta_{\mathrm{p}}$.
The blue data shows the total probability of the even-parity states and the orange data shows that of the odd-parity states.
In (a), the pump frequencies were set to $\omega_{\mathrm{p}1}/2\pi=2\times9.270$ GHz, $\omega_{\mathrm{p}2}/2\pi=2\times9.249$ GHz, $\omega_{\mathrm{p}3}/2\pi=2\times9.290$ GHz, and $\omega_{\mathrm{p}4}/2\pi=2\times9.229$ GHz
satisfying the four-body mixing condition $\omega_{\mathrm{p}1} +\omega_{\mathrm{p}2} = \omega_{\mathrm{p}3} +\omega_{\mathrm{p}4}$.
The purple, pink, and cyan vertical dashed lines show $\theta_\mathrm{p}=0$, $\pi$, and $2\pi$,
respectively, which correspond to the conditions in Fig.~\ref{fig:probs_bar}.
In (b), the pump frequencies were set to $\omega_{\mathrm{p}1}/2\pi=2\times9.270$ GHz, $\omega_{\mathrm{p}2}/2\pi=2\times9.249$ GHz, $\omega_{\mathrm{p}3}/2\pi=2\times9.289$ GHz, and $\omega_{\mathrm{p}4}/2\pi=2\times9.229$ GHz.
These pump frequencies do not meet the four-body mixing condition.
In (c), the pump frequencies were set to $\omega_{\mathrm{p}1}/2\pi=2\times9.270$ GHz, $\omega_{\mathrm{p}2}/2\pi=2\times9.250$ GHz, $\omega_{\mathrm{p}3}/2\pi=2\times9.290$ GHz, and $\omega_{\mathrm{p}4}/2\pi=2\times9.230$ GHz.
These pump frequencies satisfy $\omega_{\mathrm{p}1} +\omega_{\mathrm{p}4} = 2\omega_{\mathrm{p}2}$
and $2\omega_{\mathrm{p}1} = \omega_{\mathrm{p}2} + \omega_{\mathrm{p}3}$ in addition to the four-body mixing condition.
The distortions of the probabilities originate from the residual interactions.
}
\end{figure*}

In Fig.~\ref{fig:even_odd_prob}(a),
we show the total probabilities of the even-parity and odd-parity states
as a function of $\theta_{\mathrm{p}}$.
We swept $\theta_{\mathrm{p}1}$ in the experiment.
The data show a cosine-like dependence with a period of $4\pi$, as expected by $\breve{h}^{(4)}$.
In this experiment, the maximum (minimum) total probability
of the even-parity states was about $64.3 \pm0.2$ ($35.4 \pm0.2$) \%.
This correlation dissappears when the pump frequency condition is changed
from the four-body mixing condition [Fig.~\ref{fig:even_odd_prob}(b)]. 
This result indicates that we can achieve four-body interaction without a coupler nonlinearity.
A quantitative evaluation of the correlation will be the focus of future investigations.

We comment on the presence of residual interactions arising from KPO nonlinearities.
As shown in Fig.~\ref{fig:even_odd_prob}(c), we conducted a measurement similar to
Fig.~\ref{fig:even_odd_prob}(a),
but with different pump frequencies,
$\omega_{\mathrm{p}1}/2\pi=2\times9.270$ GHz, $\omega_{\mathrm{p}2}/2\pi=2\times9.250$ GHz,
$\omega_{\mathrm{p}3}/2\pi=2\times9.290$ GHz, and $\omega_{\mathrm{p}4}/2\pi=2\times9.230$ GHz.
These pump frequencies satisfy $\omega_{\mathrm{p}1} +\omega_{\mathrm{p}4} = 2\omega_{\mathrm{p}2}$ and
$2\omega_{\mathrm{p}1} = \omega_{\mathrm{p}2} +\omega_{\mathrm{p}3}$ in addition to $\omega_{\mathrm{p}1} +\omega_{\mathrm{p}2} = \omega_{\mathrm{p}3} +\omega_{\mathrm{p}4}$.
In this case, the residual interactions do not vanish in the rotating frame
and distort the phase dependence.
They contribute to the eigenenergy as a two-body correlation.
We found that the residual interactions in the rotating frame are expressed as $H_{\mathrm{res}1}$ in Eq.~(\ref{eq:H_res1})
with the corresponding eigenenergy $E_{\mathrm{res}1}$ in Eq.~(\ref{eq:E_res1}).
These interactions can also be derived using the unitary transformation in Eq.~(\ref{eq:Uq}).
It is important to cancel them, as they exhibit lower-order contributions than $H_{\mathrm{four}}$
in the perturbation expansion.
We predict that the interactions described in $H_{\mathrm{res}2}$ [Eq.~(\ref{eq:H_res2})] also exist,
but these interactions are higher-order terms and can be ignored.

\subsection{State control with coherent drives}\label{subsec:state-control}
We also conducted an experiment to control the state probabilities with coherent drives.
From Eq.~(\ref{eq:E1}), it is evident that varying $\theta_{\mathrm{d}j}$ enables the
controlling of the state probability of KPO $j$.

\begin{figure*}[]
\includegraphics[width=18cm]{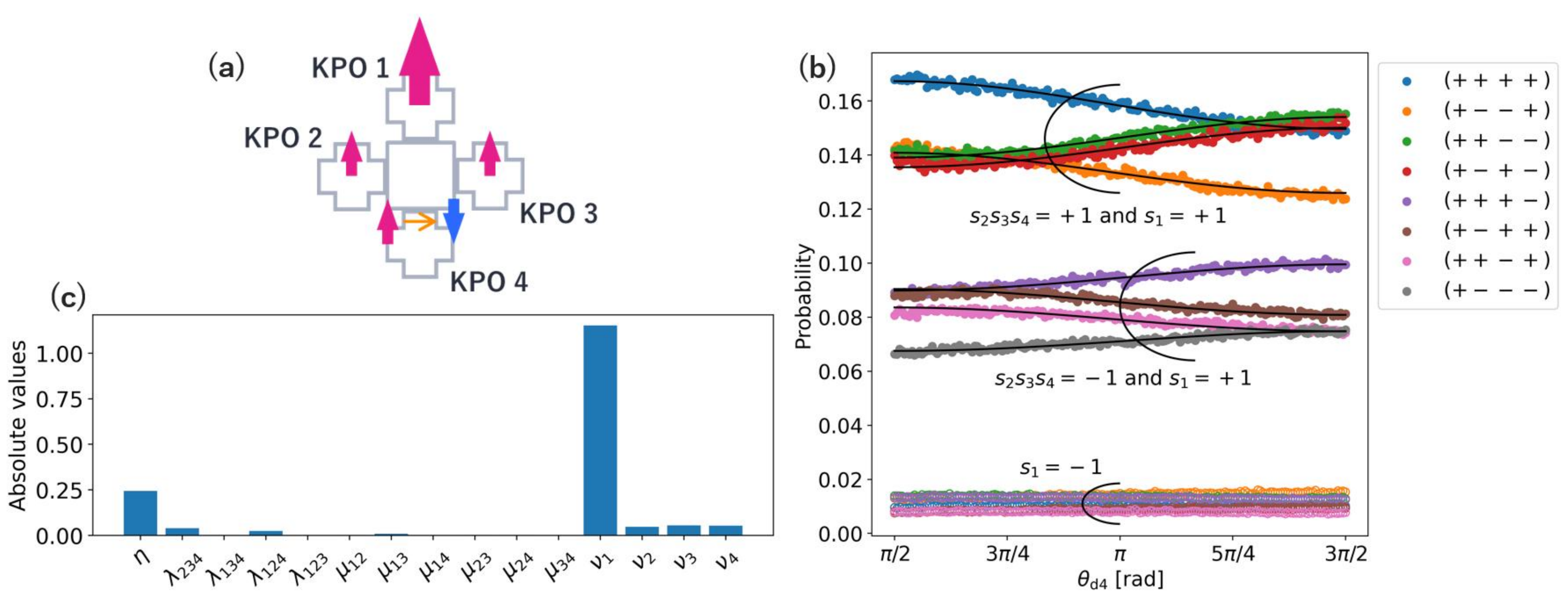}
\caption{\label{fig:LHZ_Dphase} 
  (a) Schematic of the experimental procedure.
  The arrows represent the spin states favored by the coherent drives ($s_j = \pm 1$).
  Their sizes qualitatively correspond to the amplitudes of the drives.
  We strongly apply a coherent drive for KPO 1 with $\breve{\epsilon}_1 < 0$, favoring $s_1=+1$.
  We weakly apply coherent drives for KPOs 2, 3, and 4 with $\breve{\epsilon}_2, \breve{\epsilon}_3, \breve{\epsilon}_4 <0$,
  favoring $s_2=s_3=s_4=+1$, respectively.
  We vary $\theta_{\mathrm{d}4}$ from $\pi/2$ to $3\pi/2$,
  [horizontal axis in (b)], which changes the direction of the local field applied to KPO 4,
  resulting in a flip of $s_4$ (blue arrow).
  (b) Probability of four-spin states ($s_1, s_2, s_3, s_4$) as a function of $\theta_{\mathrm{d}4}$,
  the phase of coherent drive for KPO 4.
  The legend includes only the states with $s_1=+1$.
  The black curves represent the fittings for the states with $s_1 = +1$.
  Those for the states with $s_1 = -1$ are omitted for simplicity.
  (c) Absolute values of the fitted coefficients.
}
\end{figure*}

Figure~\ref{fig:LHZ_Dphase}(a) shows the schematic of the experimental procedure.
We roughly tuned $\theta_{\mathrm{p}} = 0$ so that $\breve{h}^{(4)} < 0$, favoring $s_1 s_2 s_3 s_4 = +1$.
We strongly drove KPO 1 with $\breve{\epsilon}_1 < 0$,
favoring $s_1 = +1$ and $s_2 s_3 s_4 = +1$.
This is the implementation of a fixed qubit in the LHZ architecture~\cite{Lechner:2015bal}.
KPOs 2 and 3 were weakly driven with $\breve{\epsilon}_2 < 0$ and $\breve{\epsilon}_3 < 0$,
favoring $s_2=+1$ and $s_3=+1$, respectively.
Initially, KPO 4 was weakly driven with $\breve{\epsilon}_4 < 0$ ($\theta_{\mathrm{d}4} = \pi/2$),
favoring $s_4=+1$.
We swept the drive phase $\theta_{\mathrm{d}4}$ from $\pi/2$ to $3\pi/2$
and finally drove KPO 4 weakly with $\breve{\epsilon}_4 > 0$,
favoring $s_4=-1$.
This variation corresponds to the X-axis in Fig.~\ref{fig:LHZ_Dphase}(b),
which is described below.
We set the constant amplitudes of the drives $\epsilon_j$ for KPOs 2, 3, and 4 such that they are sufficiently weak to maintain $s_2 s_3 s_4 = +1$.
Figure~\ref{fig:LHZ_Dphase}(b) shows variation in the probability of the four-KPO states
as a function of $\theta_{\mathrm{d}4}$.
We can categorize the state probabilities into three groups.
The group exhibiting high probabilities corresponds to the four states
that satisfy $s_1 = +1$ and $s_2 s_3 s_4 = +1$, complying with the strong drive for KPO 1 with $\breve{\epsilon}_1 < 0$
and the four-body interaction with $\breve{h}^{(4)} < 0$.
The group with intermediate probabilities
corresponds to the four states that satisfy $s_1 = +1$ and $s_2 s_3 s_4 = -1$.
The hierarchical structure shows that $\breve{h}^{(4)} > \sum_{j=2}^4 |\breve{\epsilon}_j|$ is maintained
throughout the phase variation, because otherwise two groups would overlap:
the purple data would surpass the orange data at $\theta_{\mathrm{d}4}=3\pi/2$, for example.
The group with low probabilities corresponds to the eight states with $s_1 = -1$.
The probabilities of this group are very low due to the strong drive for KPO 1 with $\breve{\epsilon}_1 < 0$.

Here, we focus on the high-probability group and qualitatively consider
the dependence on $\theta_{\mathrm{d}4}$.
At $\theta_{\mathrm{d}4}=\pi/2$, the state $(s_1, s_2, s_3, s_4) = (+1, +1, +1, +1)$ 
has the highest probability (blue points) because it is favored both by the coherent drive
and the four-body interaction.
The probabilities of the other three states in the high-probability group are lower
as two of $s_2$, $s_3$, and $s_4$ are opposite to those of the state $(+1, +1, +1, +1)$.

At $\theta_{\mathrm{d}4}=3\pi/2$,
the three states in the high-probability group have similar probablities,
because one of $s_2$, $s_3$, and $s_4$ are not favored by the coherent drives in those states.
The probability of the state $(+1, -1, -1, +1)$ is relatively low (orange points) since all $s_2$, $s_3$, and $s_4$ are
disfavored by the coherent drives.

To investigate the results in quantitatively,
we assume the Boltzmann distribution for the state probabilities~\cite{Goto:boltzman,Dykman:2010sxv}.
The occurrence probability of a state $(s_1, s_2, s_3, s_4)$ is written as
\begin{equation}
  p_{s_1 s_2 s_3 s_4} = \frac{1}{Z}e^{-\beta E_{s_1 s_2 s_3 s_4}} \label{eq:boltzmann},
\end{equation}
\begin{equation}
  Z = \sum_{s_1,s_2,s_3,s_4=\pm1} e^{-\beta E_{s_1 s_2 s_3 s_4}},
\end{equation}
where $Z$ is the canonical partition function, and $\beta$ is the inverse temperature.
From the energies in Eqs.~(\ref{eq:E4})--(\ref{eq:E1}),
we assume the following effective energy:
\begin{align}
  &\beta E_{s_1 s_2 s_3 s_4} \notag \\
  &= \eta s_1 s_2 s_3 s_4 \notag\\
  &+ \lambda_{234} s_2 s_3 s_4 +\lambda_{134} s_1 s_3 s_4 +\lambda_{124} s_1 s_2 s_4 +\lambda_{123} s_1 s_2 s_3 \notag \\
  &+ \mu_{12} s_1 s_2 +\mu_{13} s_1 s_3 +\mu_{14} s_1 s_4 \notag \\
  &+\mu_{23} s_2 s_3 +\mu_{24} s_2 s_4 +\mu_{34} s_3 s_4 \notag \\
  &+ \nu_{1} s_1 +\nu_{2} s_2 +\nu_{3} s_3 +\nu_{4} s_4 \sin \theta_{\mathrm{d}4}, \label{eq:total_energy}
\end{align}
where we have the relations $\eta = \beta \breve{h}^{(4)}$, $\nu_j = \beta \breve{\epsilon}_j$ ($j= 1, 2, 3$),
and $\nu_4 = 2 \beta \epsilon_4 \alpha_4$.
We also include the three-body couplings $\lambda_{jkl}$ in the effective energy.
The data presented in Fig.~\ref{fig:LHZ_Dphase}(b) can be fitted using Eqs.~(\ref{eq:boltzmann})
and (\ref{eq:total_energy}), allowing for the extraction of the coefficients of the effective energy.
Note that the extracted coefficients include $\beta$.
For simplicity, we exclude $Z$ from the analysis
by using the following fitting functions:
\begin{equation}
  p_{s_1s_2s_3s_4} = p_{-1-1-1-1} e^{-\beta \left(E_{s_1 s_2 s_3 s_4} -E_{-1-1-1-1}\right)},
\end{equation}
\begin{equation}
  p_{-1-1-1-1} = A \exp(B\sin\theta_{\mathrm{d}4} +C),
\end{equation}
where $A$, $B$, and $C$ are the fitting parameters,
and the function of $p_{-1-1-1-1}$ is assumed from Eqs.~(\ref{eq:boltzmann}) and (\ref{eq:total_energy}).

Figure~\ref{fig:LHZ_Dphase}(c) shows the absolute values of the fitted coefficients.
The $\nu_1$ is significantly large since we strongly drove KPO 1.
As predicted from the hierarchy of the probability groups in Fig.~\ref{fig:LHZ_Dphase}(b),
the fitted parameters satisfy $|\eta| > \sum_{j=2, 3, 4}|\nu_j|$.
The $\mu_{jk}$s are nearly zero because the residual interactions,
described in Eqs.~(\ref{eq:H_res1}) and (\ref{eq:H_res2}), were eliminated in the experiment.

From the ratio between $\eta$ and $\nu_4$ obtained from the fitted results,
we roughly estimate $|h^{(4)}|/2\pi$ to be $0.6$ MHz,
which is on the same order as the value estimated from Eq.~(\ref{eq:original-h4}).
The difference between these estimations may arise from
higher-order effects of $\tilde{\omega}_j$s and $\tilde{K}_j$s.

Here, we discuss the nonzero $\lambda_{234}$ and $\lambda_{124}$.
First we consider $\lambda_{234}$.
Due to the strong drive for KPO 1,
we can consider the following unitary transformation~\cite{Xiao:2023omn}:
\begin{equation}
  h^{(4)} a_1^\dagger a_2^\dagger a_3 a_4 \to h^{(4)} a_1^\dagger a_2^\dagger a_3 a_4 + h^{(4)} \xi_1^\ast a_2^\dagger a_3 a_4\label{eq:coh_trans},
\end{equation}
where we omit the hermitian conjugates.
The $\xi_1$ is the displacement parameter for the coherent drive,
which depends on $\epsilon_1$,
and we assume the rotating-frame frequency of KPO 1 equals $\omega_{\mathrm{d}1}$.
The $\lambda_{234}$ might originate from the second term on the right-hand side of Eq.~(\ref{eq:coh_trans}).
The $\lambda_{124}$ is also nonzero
since $\epsilon_3$ was relatively large in the experiment.
When a strong coherent drive is used to prepare a fixed qubit in the LHZ scheme,
the three-body interaction may not pose a significant issue.
Even if that is not the case,
the interaction can be suppressed by appropriately weakening the drive amplitude
or adjusting the drive phase.

The experiment in Fig.~\ref{fig:LHZ_Dphase}(b) represents the first demonstration of KPO-based quantum annealing 
using the LHZ scheme by finding the minimum energy state with high success probability
under the constraint of a four-body intetaction.

\section{Conclusion}\label{sec:conclusion}
We theoretically investigated four-body interactions in KPO circuits.
We found that Kerr nonlinearities of KPOs can generate four-body interactions
without any nonlinear couplers.
Since this interaction is scalable, it can be used to implement the LHZ scheme
to embed the all-to-all two-body interactions of logical spins in a planar physical qubit network.
Therefore, we expect that the fabrication and operation
of the Ising machine based on the LHZ scheme will become much simpler
by using the new circuits.
Numerical calculations show that KPO nonlinearities can generate large magnitudes of four-body couplings
comparable to that originating from a transmon-like coupler.

We designed and fabricated a device incorporating four capacitively coupled KPOs on the basis of our theoretical investigation.
Using this device, we prepared the oscillation state for each KPO
such that their pump frequencies satisfy the four-body mixing condition.
The observed correlation between the oscillation states confirms the presence of a four-body interaction.
It is important to choose proper pump frequencies to cancel undesired residual interactions,
which manifest themselves as distortion in the correlation curve for a four-body interaction.
We also conducted an experiment to control the state probabilities using coherent drives to demonstrate quantum annealing with the LHZ scheme.
In the experiment, the states favored by the four-body interaction exhibited higher probabilities than the other states
indicating the presence of
a relatively strong four-body interaction compared with the coherent drives.
A rough estimation indicates that the experimental results are in agreement with
the theoretical prediction for the magnitude of $h^{(4)}$.
A precise evaluation is left for future work.

The LHZ scheme can be applied not only to quantum annealing but also
to other areas of quantum computing,
such as quantum walks~\cite{Bennett:2025yst},
the quantum approximate optimization algorithm (QAOA)~\cite{Weidinger:2024ipw},
and quantum error correction~\cite{Messinger:2024vhl,Tiurev:2025gbv}.
The circuit demonstrated in this study
can also be used for the implementation of those proposals.

\section*{ACKNOWLEDGMENTS}
We thank R.~Miyazaki, Y.~Susa, and K.~Ishihara for valuable discussions,
Y.~Kitagawa for assistance with device fabrication,
and M.~Nagasaku for help with the experimental setup.
The devices
were fabricated in the Superconducting Quantum Circuit Fabrication Facility (Qufab)
in National Institute of Advanced Industrial Science and Technology (AIST).
This paper is based on results obtained
from Project No. JPNP16007 commissioned by the New
Energy and Industrial Technology Development Organization (NEDO), Japan.

\section*{Appendix A: Hamiltonian of conventional unit circuit}
In this appendix, we derive the effective Hamiltonian of the circuit in Fig.~\ref{fig:original},
but we set $L_g = L_q = 0$ for simplicity.
The circuit contains four KPOs and a nonlinear coupler.
The Hamiltonian is written as
\begin{equation}
  H = T \left( \bm{Q} \right) +V \left( \bm{\Phi}\right) = \frac{1}{2} \bm{Q}^T \bm{C}^{-1} \bm{Q} +V (\bm{\Phi}),
\end{equation}
\begin{equation}
  \bm{Q}^T = \left( Q_1, \, Q_2, \, Q_3, \, Q_4, \, Q_5, \, Q_6 \right),
\end{equation}
\begin{equation}
  \bm{\Phi}^T = \left( \Phi_1, \, \Phi_2, \, \Phi_3, \, \Phi_4, \, \Phi_5, \, \Phi_6 \right),
\end{equation}
where $Q_k$ and $\Phi_k$ are the charge and flux for node $k$, respectively.
The term $T \left( \bm{Q} \right)$ is the kinetic term and $V \left(\bm{\Phi} \right)$ is the potential term.
The $\bm{C}$ is the capacitance matrix, which is written as
\begin{widetext}
  \begin{equation}
    \bm{C} = \begin{pmatrix} C_q +C_c & 0 & 0 & 0 & -C_c & 0 \\
      0 & C_q +C_c & 0 & 0 & -C_c & 0 \\
      0 & 0 & C_q +C_c & 0 & 0 & -C_c \\
      0 & 0 & 0 & C_q +C_c & 0 & -C_c \\
      -C_c & -C_c & 0 & 0 & C_g +2C_c & -C_g \\
      0 & 0 & -C_c & -C_c & -C_g & C_g +2C_c
    \end{pmatrix}.
  \end{equation}
\end{widetext}
The inverse matrix has seven independent elements
[their approximate expressions are given in Eqs.~(\ref{eq:g12_g13})--(\ref{eq:tCg})]:
  \begin{equation}
    \bm{G} = \bm{C}^{-1} = \begin{pmatrix} G_{11} & G_{12} & G_{13} & G_{13} & G_{15} & G_{16} \\
      G_{12} & G_{11} & G_{13} & G_{13} & G_{15} & G_{16} \\
      G_{13} & G_{13} & G_{11} & G_{12} & G_{16} & G_{15} \\
      G_{13} & G_{13} & G_{12} & G_{11} & G_{16} & G_{15} \\
      G_{15} & G_{15} & G_{16} & G_{16} & G_{55} & G_{56} \\
      G_{16} & G_{16} & G_{15} & G_{15} & G_{56} & G_{55}
    \end{pmatrix}.
  \end{equation}
The kinetic term is written as
\begin{align}
  T \left(\bm{Q} \right) = &\frac{1}{2} \sum_j^4 G_{11} Q_j^2 +\frac{1}{2} G_{55} \left( Q_5^2 +Q_6^2\right) \notag \\
  &+G_{12} \left( Q_1 Q_2 +Q_3 Q_4\right) \notag \\
  &+G_{13} \left( Q_1 +Q_2\right) \left( Q_3 +Q_4\right) \notag \\
  &+\left( Q_1 +Q_2\right) \left( G_{15} Q_5 +G_{16} Q_6\right) \notag \\
  &+\left( Q_3 +Q_4\right) \left( G_{16} Q_5 +G_{15} Q_6\right) \notag \\
  &+G_{56} Q_5 Q_6.
\end{align}
By defining the new degrees of freedom of branch charges and capacitance parameters as
\begin{equation}
  Q_{\pm} = \frac{1}{\sqrt{2}} \left( Q_5 \pm Q_6\right),
\end{equation}
\begin{equation}
  G_{\pm} = \frac{1}{\sqrt{2}} \left( G_{15} \pm G_{16}\right),
\end{equation}
\begin{equation}
  C_{\pm} = \frac{1}{G_{55} \pm G_{56}},
\end{equation}
we obtain
\begin{align}
  T \left( \bm{Q} \right) = &\sum_j^4 \frac{Q_j^2}{2\tilde{C}_q} +\frac{Q_+^2}{2C_+} +\frac{Q_-^2}{2C_-} \notag \\
  &+G_{12} \left( Q_1 Q_2 +Q_3 Q_4\right) \notag \\
  &+G_{13} \left( Q_1 +Q_2 \right) \left( Q_3 +Q_4\right) \notag \\
  &+G_+Q_+ \sum_j^4 Q_j + G_-Q_-\sum_j^4 s_j Q_j,
\end{align}
where $\tilde{C}_q = 1/G_{11}$ is the effective capacitance of a KPO.
Note when $C_q \sim C_g \gg C_c$, $\tilde{C}_q \simeq C_q$.
The $C_-$ is of the same order as $C_g$ ($C_- \simeq 2C_g$), while $C_+$ is comparable to $C_c$ ($C_+ \simeq 2C_c$);
thus, we regard $Q_-$ as the coupler charge.
Hereafter, we neglect the mode associated with $Q_+$, as its corresponding frequency is much higher
than the KPO and coupler frequencies.

Next we consider the potential term $V \left( \bm{\Phi} \right)$, which is written as
\begin{equation}
  V \left( \bm{\Phi}\right) = \sum_j^4 V_q (\Phi_j) +V_g \left( \Phi_-\right),
\end{equation}
\begin{equation}
  \Phi_- = \frac{1}{\sqrt{2}} \left( \Phi_5 -\Phi_6\right),
\end{equation}
where $V_q \left( \Phi_j\right)$ represents the potential of KPO $j$ and $V_g \left( \Phi_-\right)$ represents that of the coupler. 
The $\Phi_-$ is regarded as the coupler flux.
When we use SQUIDs for the KPOs and coupler, their potentials are written as
\begin{equation}
  V_q \left( \varphi_j\right) = -E_{Jj} \cos \varphi_j,
\end{equation}
\begin{equation}
  V_g \left( \varphi_-\right) = -E_{Jg} \cos \left( \sqrt{2} \varphi_-\right),
\end{equation}
where $E_{Jj}$ is the Josephson energy of KPO $j$, and $E_{Jg}$ is that of the coupler.
The $\varphi_j$ and $\varphi_-$ are defined as
$\varphi_j = \Phi_j/\varphi_0$ and $\varphi_- = \Phi_-/\varphi_0$, respectively.
We expand the potentials to the fourth order $\cos x = 1 -x^2/2! +x^4/4!$.

The Hamiltonian is quantized with:
\begin{equation}
  n_j = \frac{Q_j}{2e}, \qquad n_- = \frac{Q_-}{2e},
\end{equation}
\begin{equation}
  \varphi_j \to \varphi_{Zj} \left( a_j +a_j^\dagger\right), \qquad n_j \to -\frac{i}{2\varphi_{Zj}} \left( a_j -a_j^\dagger\right),
\end{equation}
\begin{equation}
  \varphi_- \to \varphi_{Zg} \left( a_g +a_g^\dagger\right), \qquad n_- \to -\frac{i}{2\varphi_{Zg}} \left( a_g -a_g^\dagger\right),
\end{equation}
where they satisfy $\left[ \varphi_j, \, n_j\right] = i$ and $\left[ \varphi_-, \, n_-\right] = i$;
$\left[ a_j, \, a_j^\dagger\right] = 1$, and $\left[ a_g, \, a_g^\dagger \right] = 1$.
The $\varphi_{Zj}$ and $\varphi_{Zg}$ are the renormalization constants, which are written as
\begin{equation}
  \varphi_{Zj} = \left( \frac{e^2 L_{Jj}}{\varphi_0^2 \tilde{C}_q}\right)^{\frac{1}{4}}, \qquad \varphi_{Zg} = \left( \frac{e^2 L_{Jg}}{4 \varphi_0^2 \tilde{C}_g}\right)^{\frac{1}{4}}.
\end{equation}
Here, $\tilde{C}_g = C_-/2$ is the effective capacitance of the coupler, and $L_{Jj} = \varphi_0^2/E_{Jj}$ and $L_{Jg} = \varphi_0^2/E_{Jg}$ are the Josephson inductances of KPO $j$ and the coupler, respectively.
Finally, we obtain the following quantized Hamiltonian:
\begin{align}
  H/\hbar&= \sum_j^4 \left\{ \omega_j a_j^\dagger a_j -\frac{K_j}{12} \left( a_j +a_j^\dagger\right)^4\right\}\notag \\
  &+\omega_g a_g^\dagger a_g -\frac{K_g}{12} \left( a_g +a_g^\dagger\right)^4 \notag \\
  &-\sum_{j < k} h_{jk} \left( a_j -a_j^\dagger\right) \left( a_k -a_k^\dagger\right) \notag \\
  &-\sum_j^4 s_j g_{j} \left( a_j -a_j^\dagger\right) \left( a_g -a_g^\dagger\right),
\end{align}
\begin{equation}
  K_j = \frac{e^2}{2\tilde{C}_q}, \qquad K_g = \frac{e^2}{2\tilde{C}_g},
\end{equation}
\begin{equation}
  \omega_j = \frac{1}{\sqrt{L_{Jj} \tilde{C}_q}}, \qquad \omega_g = \frac{1}{\sqrt{L_{Jg} \tilde{C}_g}},
\end{equation}
\begin{equation}
  h_{jk}\notag = \begin{cases}  \frac{1}{2} G_{12} \tilde{C}_q \sqrt{\omega_j \omega_k}, \quad &\left( j, \, k\right) \in \left\{ \left( 1, \, 2\right), \, \left( 3, \, 4\right)\right\}, \\
  \frac{1}{2} G_{13} \tilde{C}_q \sqrt{\omega_j \omega_k}, \quad &\mathrm{otherwise} \left(j\neq k\right),
\end{cases}
\end{equation}
\begin{equation}
  g_{j} = \frac{G_-}{\sqrt{2}} \sqrt{\tilde{C}_q \tilde{C}_g \omega_j \omega_g}.
\end{equation}
We consider the coupling constants $h_{jk}$ and $g_{j}$.
With the realistic conditions of capacitances $C_q \sim C_g \gg C_c$,
we approximately obtain
\begin{equation}
  G_{12} \sim G_{13} \simeq \frac{C_c}{4 C_q^2} \label{eq:g12_g13},
\end{equation}
\begin{equation}
  G_- = \frac{1}{\sqrt{2}} \left( G_{15} -G_{16}\right) \simeq \frac{1}{2 \sqrt{2}} \frac{C_c}{C_q C_g},
\end{equation}
\begin{equation}
  \tilde{C}_q = \frac{1}{G_{11}} \simeq C_q, \label{eq:tCq}
\end{equation}
\begin{equation}
  \tilde{C}_g = \frac{C_-}{2} = \frac{1}{2} \left( G_{55} -G_{56}\right)^{-1} \simeq C_g, \label{eq:tCg}
\end{equation}
then we obtain
\begin{equation}
  h_{jk} \simeq \frac{C_c}{8 C_q} \sqrt{\omega_j \omega_k},
\end{equation}
\begin{equation}
  g_{j} \simeq \frac{C_c}{4\sqrt{C_q C_g}} \sqrt{\omega_j \omega_g}.
\end{equation}
These expressions are used in the numerical study of the four-body couplings.

\section*{Appendix B: unitary transformation to derive four-body interaction} \label{sec:app-B}
We introduce a unitary transformation in the dispersive regime
following previous studies~\cite{Hajr:2024eme, Blais:2020wjs}.
Consider the following coupled resonators' Hamiltonian:
\begin{equation}
  H/\hbar = \omega_a a^\dagger a +\omega_b b^\dagger b +g \left( a^\dagger b +a b^\dagger\right),
\end{equation}
where $a$ ($b$) and $\omega_a$ ($\omega_b$) are the annihilation operator and resonance frequency of resonator A (B), respectively,
and $g$ is their coupling constant.
We define the following unitary operator:
\begin{equation}
  U = e^{\Lambda \left( a^\dagger b -a b^\dagger\right)},
\end{equation}
where $\Lambda \ll 1$ is the perturbative parameter.
Under the first-order transformation, the annihilation operators transform as
\begin{equation}
  a \to a +\Lambda b,
\end{equation}
\begin{equation}
  b \to b -\Lambda a.
\end{equation}
We then obtain the following effective Hamiltonian:
\begin{equation}
  U^\dagger H U/\hbar \overset{\mathrm{1st}}{=} H/\hbar +\Lambda \left( \omega_a -\omega_b\right) \left( ab^\dagger +a^\dagger b\right) +2\Lambda g \left( -a^\dagger a +b^\dagger b\right).
\end{equation}
We choose $\Lambda = -g/\left( \omega_a -\omega_b\right)$, thus can approximately diagonalize the Hamiltonian.

To derive the four-body interactions, we defined the following unitary transformations in the main text:
\begin{equation}
  U_g = \mathrm{exp} \left[ \sum_j \left\{ -s_j \tilde{g}_{j} \left( a_j^\dagger a_g -a_j a_g^\dagger \right)\right\} \right], \quad \tilde{g}_{j} = \frac{g_{j}}{\omega_j -\omega_g},
\end{equation}
\begin{equation}
  U_q = \mathrm{exp} \left[ \sum_{j<k} \left\{ -\tilde{h}_{jk} \left( a_j^\dagger a_k -a_j a_k^\dagger \right)\right\} \right], \quad \tilde{h}_{jk} = \frac{h_{jk}}{\omega_j -\omega_k} \label{eq:app-Uq},
\end{equation}
where $s_j = 1$ ($j = 1, \, 2$) and $s_j = -1$ ($j = 3, \, 4$).

We derive the four-body coupling $g^{(4)}$ by using $U_g$.
The $a_g$ transforms at the first order as follows:
\begin{equation}
  a_g \to a_g -\sum_j s_j \tilde{g}_{j} a_j.
\end{equation}
The Kerr nonlinearity of the coupler can be transformed as
\begin{align}
  &-\frac{K_g}{2} a_g^{\dagger 2} a_g^2 \notag \\
  &\to -\frac{K_g}{2} \left( a_g^\dagger -\sum_j s_j \tilde{g}_{j} a_j^\dagger \right)^2 \left( a_g^\dagger -\sum_j s_j \tilde{g}_{j} a_j^\dagger \right)^2 \notag \\
  &\to -2K_g\tilde{g}_{1} \tilde{g}_{2} \tilde{g}_{3} \tilde{g}_{4} a_1^\dagger a_2^\dagger a_3 a_4 \notag \\
  &\equiv-g^{(4)}a_1^\dagger a_2^\dagger a_3 a_4\label{eq:coupler_kerr_trans}.
\end{align}

In a similar manner, we can obtain the four-body coupling $h^{(4)}$ by using $U_q$.
The annihilation operator $a_j$ transforms at the first order as
\begin{equation}
  a_j \to a_j -\sum_{j\neq k}\tilde{h}_{jk} a_k = a_j +\sum_{j\neq k} \tilde{h}_{kj} a_k,
\end{equation}
where $\tilde{h}_{kj} = -\tilde{h}_{jk}$.
Consider the Kerr nonlinearity of KPO 1, which transforms as
\begin{align}
&-\frac{K_1}{2} a_1^{\dagger2} a_1^2 \notag \\
&\to -\frac{K_1}{2} \left( a_1^\dagger +\sum_{1\neq j} \tilde{h}_{j1} a_j^\dagger\right)^2 \left( a_1 +\sum_{1\neq j} \tilde{h}_{j1} a_j \right)^2 \notag \\
&\to -2K_1 \tilde{h}_{21} \tilde{h}_{31} \tilde{h}_{41} a_1^\dagger a_2^\dagger a_3 a_4\label{eq:kpo_kerr_trans}.
\end{align}
Taking the other KPO nonlinearities into account, we obtain four-body coupling $h^{(4)}$ as in Eq.~(\ref{eq:original_h4}).

The perturbative expressions for $\tilde{\omega}_j$ and $\tilde{K}_j$ in
Eq.~(\ref{eq:H_KPO}) can be
obtained by applying $U_q$.
The $\tilde{\omega}_j$ up to $\mathcal{O} \left(\Lambda^2 \right)$
is expressed as
\begin{equation}
  \tilde{\omega}_j = \omega_j -K_j +\Omega_j +\mathcal{O} \left(\Lambda^3 \right)\label{eq:tomega},
\end{equation}
\begin{equation}
  \Omega_j = \sum_{k\neq j} h_{jk}\tilde{h}_{kj} +\sum_{k, l\neq j} h_{jk}\tilde{h}_{kl}\tilde{h}_{lj} -\sum_{\substack{k, l\neq j\\k<l}} \tilde{h}_{jk}h_{kl}\tilde{h}_{lj},
\end{equation}
where $-K_j$ in Eq.~(\ref{eq:tomega}) is the Lamb shift.
The $\tilde{K}_j$ is written as
\begin{equation}
  \tilde{K}_j = \left(1 +2\sum_{k\neq j} \tilde{h}_{jk} \tilde{h}_{kj} \right) K_j +\mathcal{O} \left( \Lambda^3\right)\label{eq:tK}.
\end{equation}

\section*{Appendix C: parameters of SNAIL KPOs} \label{sec:app-C}
\begin{figure}[tb]
\begin{center}
\includegraphics[width=8cm]{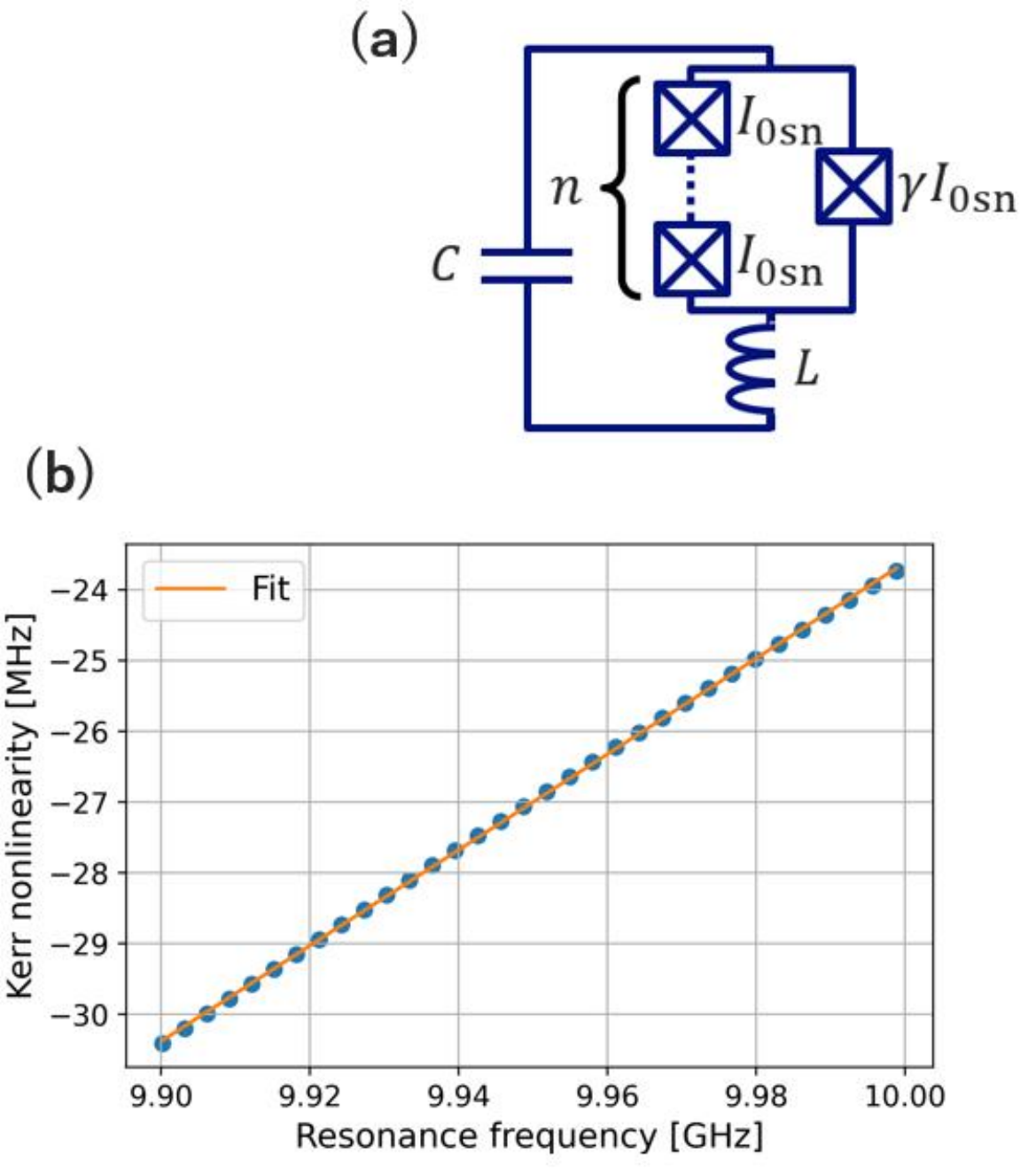}
\caption{
(a) Schematic of a SNAIL KPO circuit.
The $C$ and $L$ are the geometric capacitance and inductance, respectively.
The $I_{0\mathrm{sn}}$ is the critical current of a single junction in the left branch of
the superconducting loop of the SNAIL.
The $n$ is the number of junctions.
The $\gamma$ represents the ratio of the critical currents of the Josephson junctions
that satisfies $0 < \gamma < 1$.
(b) Relation between the Kerr nonlinearity and resonance frequency of the SNAIL.
The frequency control is achieved by applying an external flux around $\varphi_X/2\pi\approx 0.47$.
We used $C = 200$ fF, $L = 100$ pH, $I_{0\mathrm{sn}} = 1250$ nA, $n = 2$, and $\gamma = 0.3$.
To generate the plots, we numerically solve Eq.~(\ref{eq:SNAIL_current}) to determine $\bar{\varphi}$ and $c_k$s,
and then calculate $K$ and $\omega$ using Eqs.~(\ref{eq:SNAIL_kerr}) and (\ref{eq:SNAIL_freq}), respectively.
}
\label{fig:SNAIL_KPO}
\end{center}
\end{figure}

We investigate a Kerr nonlinearity of the SNAIL KPO following Frattini et al.~\cite{Frattini:2018ahg}.
The circuit structure is shown in Fig.~\ref{fig:SNAIL_KPO}(a).
In the figure, $I_{0\mathrm{sn}}$ is the critical current of the junctions
and $\gamma$ is a parameter satisfying $0 < \gamma < 1$.
The potential of the SNAIL is written as
\begin{equation}
  U \left( \varphi, \, \varphi_X \right) = -\gamma \varphi_0 I_{0\mathrm{sn}} \cos \varphi -n \varphi_0 I_{0\mathrm{sn}} \cos \left( \frac{\varphi_X -\varphi}{n}\right),
\end{equation}
where $\varphi$ is the superconducting phase across the junction with $\gamma I_{0\mathrm{sn}}$,
and $\varphi_X$ is the phase of the applied flux.
We need to find $\bar{\varphi}$ satisfying the following condition:
\begin{equation}
  I \left( \bar{\varphi}, \, \varphi_X \right) = \gamma I_{0\mathrm{sn}} \sin \bar{\varphi} -n I_{0\mathrm{sn}} \sin \left( \frac{\varphi_X -\bar{\varphi}}{n}\right) = 0 \label{eq:SNAIL_current},
\end{equation}
and we define the expansion coefficients as
\begin{equation}
  c_k = \left. \frac{1}{\varphi_0 I_{0\mathrm{sn}}} \frac{d^k U}{d \varphi^k} \right|_{\varphi=\bar{\varphi}} \label{eq:SNAIL_ck}.
\end{equation}
The Kerr nonlinearity of the SNAIL KPO is written as
\begin{equation}
  \hbar K = -\frac{p^3}{c_2} \left[ c_4 -\frac{3c_3^2}{c_2} \left( 1 -p\right) -\frac{5}{3} \frac{c_3^2}{c_2} p\right] \frac{e^2}{2C} \label{eq:SNAIL_kerr},
\end{equation}
where
\begin{equation}
  p = \frac{\varphi_0}{\varphi_0 +c_2 L I_{0\mathrm{sn}}}.
\end{equation}
The resonance frequency is written as
\begin{equation}
  \omega = \frac{1}{\sqrt{C \left( L +\frac{\varphi_0}{c_2 I_{0\mathrm{sn}}}\right)}}\label{eq:SNAIL_freq}.
\end{equation}

We choose $I_{0\mathrm{sn}} = 1250$ nA and $\gamma = 0.3$ for the SNAIL to obtain negative Kerr nonlinearity
(the nonlinearity of a single junction is positive in this context).
Figure~\ref{fig:SNAIL_KPO}(b) shows the relation between the resonance frequency and Kerr nonlinearity of the SNAIL KPO.
The nonlinearity is approximately proportional to the resonance frequency in this frequency range.
We can then obtain a one-to-one correspondence between them from the fit (orange curve).

\section*{Appendix D: circuit designs for cancellation of cross-Kerr interactions}
While not addressed in the main text,
in the circuit shown in Fig.~\ref{fig:original},
there exist cross-Kerr interactions $\chi_{jg}a_j^\dagger a_j a_g^\dagger a_g$ between KPOs and the coupler,
as well as the cross-Kerr interaction $\chi_{jk} a_j^\dagger a_j a_k^\dagger a_k$ ($k\neq j$) between KPOs
with $\chi_{jg}$ and $\chi_{jk}$ being the respective coupling constants.
They remain in the rotating frame under a four-body mixing condition.
According to Ref.~\cite{Puri:2017hqg}, cross-Kerr couplings do not lead to
significant errors in quantum annealing,
but as a precaution, we introduce ideas to cancel the cross-Kerr interactions
in this appendix.
Since a cross-Kerr interaction is basically proportional to the sum of 
Kerr nonlinearities as explained below, the interaction can be canceled using both
positive and negative Kerr nonlinearities.

\begin{figure*}[tb]
\includegraphics[width=18cm]{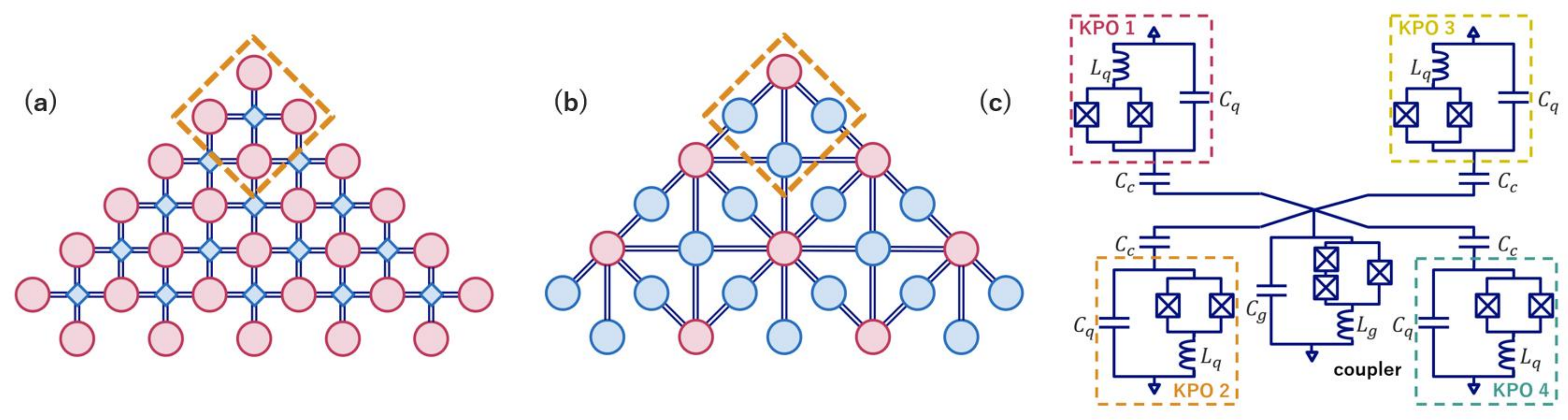}
\caption{\label{fig:cancel_cross_kerr} 
  Arrangement of KPOs and couplers for cancellation of cross-Kerr interactions.
  (a) Arrangement for the unit circuit with $g^{(4)}$.
  The red circles represent KPOs with positive nonlinearities,
  and the blue diamonds represent couplers,
  which are grounded and have negative nonlinearities.
  (b) Arrangement for the unit circuit with four-body coupling $\tilde{h}^{(4)}$.
  The red (blue) circles represent KPOs with positive (negative) nonlinearities.
  In (a) and (b), The double wires represent connecting wires interrupted by a coupling capacitance.
  The regions outlined by the orange dashed squares represent unit cells.
  (c) An example of the unit circuit for (a).
  We use a SNAIL as the coupler with a negative nonlinearity.
}
\end{figure*}

For the circuit using a nonlinear coupler,
we propose the circuit in Fig.~\ref{fig:cancel_cross_kerr}(a).
An example of the unit circuit is shown in Fig.~\ref{fig:cancel_cross_kerr}(c),
where the coupler is grounded.
As mentioned in Sec.~\ref{sec:locality},
the two-body interactions between KPOs are suppressed in this case,
although there are cross-Kerr interactions between KPOs and couplers.
From the transformation shown in Eq.~(\ref{eq:coupler_kerr_trans}),
we obtain $-2\tilde{g}_j^2K_g a_j^\dagger a_j a_g^\dagger a_g$ at the leading order.
On the other hand, we can also consider the following transformation:
\begin{align}
  &-\frac{K_j}{2} a_j^{\dagger2} a_j^2 \notag \\
  &\to -\frac{K_j}{2} \left( a_j^\dagger -s_j \tilde{g}_j a_g^\dagger\right)^2 \left( a_j -s_j \tilde{g}_j a_g\right)^2\notag \\
  &\to-2\tilde{g}_j^2K_j a_j^\dagger a_j a_g^\dagger a_g.
\end{align}
Therefore, $\chi_{jg} = -2\tilde{g}_j^2 \left(K_j +K_g\right)$ is obtained as the cross-Kerr coupling.
To suppress the interactions,
we set the Kerr nonlinearities to satisfy
$\left|K_j\right| = -\left|K_g\right|$ ($j=1,..., 4$) in each unit circuit.
For example, we can use SQUID KPOs and a SNAIL coupler in the unit circuit
as the coupler exhibits negative nonlinearity.

Without nonlinear couplers, we propose the circuit as shown in Fig.~\ref{fig:cancel_cross_kerr}(b).
We use the unit circuit with the four-body coupling $\tilde{h}^{(4)}$.
Neglecting the interactions between KPOs not connected with double lines,
there are three cross-Kerr interactions between KPOs in a unit circuit.
In the circuit shown in Fig.~\ref{fig:circuits_qubit_4bi}(b),
which is the unit cell of the circuit shown in Fig.~\ref{fig:cancel_cross_kerr}(b),
the cross-Kerr interactions involving KPO 4
with KPOs 1, 2, and 3
are expressed as $\chi_{j4}=-2\tilde{h}_{j4}^2 \left(K_j +K_4\right)$ ($j=1,..., 3$),
which are derived from the trasformation of the KPO nonlinearities, as shown in Eq.~(\ref{eq:kpo_kerr_trans}) for KPO 1.
We can thus choose the Kerr nonlinearities to be $\left|K_j\right| = -\left|K_4\right|$ ($j=1,..., 3$).
Other cross-Kerr interactions in the circuit are suppressed at the leading order.

\section*{Appendix E: device fabrication and experimental setup}
The device was fabricated on a high-resistivity Si substrate with a thickness of $380\,\mathrm{\mu}$m.
All the circuits, except for the Josephson junctions and airbridges,
were formed by a 100-nm-thick sputtered Nb film, 
which underwent dry etching using $\mathrm{CF_4}$ gas.
The Josephson junctions were fabricated in a separate lithographic process by 
shadow evaporation of Al, preceded by Ar-ion milling to eliminate surface oxides of the Nb film.

The device chip was housed in a magnetic shield and cooled to temperatures below 15 mK using
a dilution refrigerator.

We summarize the room-temperature electronics for the time-domain measurement.
A local oscillator, Keysight M9347A DDS, provides the system with continuous-wave microwaves
with a frequency of $9.3$ GHz.
We used arbitrary waveform generators, Keysight M3202A, with a sampling rate of
1 GSa/s for the baseband signal to generate the pump drives and coherent drives.
The output signals from the KPOs were recorded by an analog-to-digital converter,
Keysight M3102A with a sampling rate of 500 MSa/s.
See Appendix B of Ref.~\cite{Yamaji:2022ikh} for details.


\end{document}